\documentclass[fleqn,usenatbib]{mnras}

\usepackage{amssymb}

\usepackage{newtxtext,newtxmath}
\usepackage{xcolor}

\usepackage[T1]{fontenc}
\usepackage{gensymb}

\DeclareRobustCommand{\UD}[3]{#2}
\let\UDthebibliography\thebibliography
\def\thebibliography{\DeclareRobustCommand{\UD}[3]{##3}\UDthebibliography}

\usepackage{graphicx}	
\usepackage{amsmath}	
\usepackage{amssymb}	
\usepackage{pdflscape}	
\usepackage{hyperref}

\newcommand{\cpd}{CPD-62$^{\circ}$\,2717}
\newcommand{\bmag}{$\langle B\rangle$}

\newcommand{\blo}{$\langle B\rangle_{\rm low}$}
\newcommand{\bhi}{$\langle B\rangle_{\rm hi}$}

\title[{\cpd}]{Confident Detection of Doubly-Ionized Thorium in the Extreme Ap Star {\cpd}}

\author[S. Drew Chojnowski et al.]{
S. Drew Chojnowski$^{1}$,\thanks{E-mail: stephen.chojnowski@montana.edu}
Swetlana Hubrig$^{2}$,
David L. Nidever$^{1}$,
Ewa Niemczura$^{3}$,
Jonathan Labadie-Bartz$^{4}$,\newauthor
Gautier Mathys$^{5}$,
Sten Hasselquist$^{6}$
\\
$^{1}$Department of Physics, Montana State University, P.O. Box 173840, Bozeman, MT 59717-3840 \\
$^{2}$Leibniz-Institut f\"{u}r Astrophysik Potsdam (AIP), An der Sternwarte 16, 14482 Potsdam, Germany\\
$^{3}$Instytut Astronomiczny, Uniwersytet Wroc\l{}awski, Kopernika 11, 51-622 Wroc\l{}aw, Poland \\
$^{4}$LESIA, Paris Observatory, PSL University, CNRS, Sorbonne Universit\'{e}, Universit\'{e} de Paris, 5 place Jules Janssen, 92195 Meudon, France\\
$^{5}$European Southern Observatory, Alonso de Cordova 3107, Vitacura, Santiago, Chile\\
$^{6}$Space Telescope Science Institute, 3700 San Martin Drive, Baltimore, MD 21218, USA
}

\date{Accepted XXX. Received YYY; in original form ZZZ}

\pubyear{2023}

\begin{document}
\label{firstpage}
\pagerange{\pageref{firstpage}--\pageref{lastpage}}
\maketitle

\begin{abstract}
Despite the universe containing primordial thorium (Th) of sufficient abundance to appear in stellar spectra, detection of Th has to date been tentative and based on just a few weak and blended lines. Here, we present convincing evidence not only for the first Th detection in a magnetic chemically peculiar Ap star but also for the first detection of Th~{\sc iii} in a stellar spectrum. {\cpd} was initially recognized as a highly-magnetized Ap star thanks to resolved magnetically split lines captured in $H$-band spectra from the SDSS/APOGEE survey. The star was subsequently pinpointed as extraordinarily peculiar when careful inspection of the $H$-band line content revealed the presence of five lines of Th~{\sc iii}, none of which are detected in the other $\sim1500$ APOGEE-observed Ap stars. Follow-up with the VLT+UVES confirmed a similarly peculiar optical spectrum featuring dozens of Th~{\sc iii} lines, among other peculiarities. Unlike past claims of Th detection, and owing to high-resolution observations of the strong ($\sim$8--12\,kG) magnetic field of {\cpd}, the detection of Th~{\sc iii} can in this case be supported by matches between the observed and theoretical magnetic splitting patterns. Comparison of {\cpd} to stars for which Th overabundances have been previously reported (e.g., Przybylski's Star) indicate that only for {\cpd} is the Th detection certain. Along with the focus on Th~{\sc iii}, we use time series measurements of the magnetic field modulus to constrain the rotation period of {\cpd} to $\sim$4.8 years, thus establishing it as a new example of a super-slowly-rotating Ap star. 
\end{abstract}

\begin{keywords}
stars: magnetic fields -- stars: chemically peculiar
\end{keywords}



\section{Introduction} \label{intro}
Considering the unknown origin of super-strong magnetic fields in stars that are expected to have mostly if not fully radiative envelopes, there is no such thing as a `normal' Ap star. However, since being defined more than 60 years ago \citep{babcock1958} as a distinct sub-group of slowly rotating, highly magnetized, mostly A-type stars whose spectra show chemical peculiarities when compared to their non-magnetic analogues, a few examples have stood out as particularly peculiar and extreme. Perhaps the best-known and most peculiar Ap star (if not the most peculiar star in general) is HD\,101065, also known as Przybylski's Star. It was named after its discoverer \citep{przybylski1961}, and will be referred to as ``PS'' throughout this work.

In his initial studies of optical spectra of PS, Przybylski struggled to identify any light elements aside from hydrogen and calcium, instead finding only a few lines of strontium and barium and literally thousands of lines from rare earth elements \citep[or REE; $57\geq Z\geq71$;][]{przybylski1963a}. Even though the vast majority of Ap stars exhibit surface underabundances of light elements like helium and often extreme overabundances of heavier elements, the fact that elements like iron seemed to be missing altogether from spectra of PS left astronomers baffled for years \citep{przybylski1977}. Lines from gold have yet to be confidently identified in spectra of PS and might never be owing to the pileup of strong lines from singly-ionized REE around the wavelengths of the known optical Au~{\sc ii} lines, but PS can nonetheless be considered a spectroscopic goldmine. Remarkably, the presence of the REE holmium was confirmed in the atmosphere of PS before it had been confirmed in any other star including the Sun \citep{przybylski1963b}. 

Whereas the surface abundances of PS are undoubtedly extreme, the associated 2.3\,kG global magnetic field \citep{wolff1976,cowley2000} of PS is certainly not given that Ap star magnetic fields can reach extraordinary strengths of up to 34\,kG \citep[in the case of Babcock's Star;][]{1960ApJ...132..521B}, with numerous examples of $>$10\,kG \citep{mathys2017, giarrusso2022, choj2019}. On the other hand, the effective temperature of PS \citep[$T_{\rm eff}\approx6500$\,K, hence Fp being more accurate than Ap;][]{cowley2000,shulyak2010} places it at the extreme cool end of the known $T_{\rm eff}$ range for Ap stars and the possible 188 year rotation period \citep[][]{hubrig2018} places PS as the most extreme member of the super-slowly rotating Ap stars \citep[ssrAp;][]{mathys2022}. In addition to the spectroscopic peculiarities, PS is photometrically variable due to non-radial pulsation with a $\sim$12\,minute period \citep{kurtz1979}. PS was the first known example of what are now referred to as rapidly oscillating Ap stars \citep[or roAp;][]{kurtz1982}.

Motivated by knowledge of the long-lived isotopes of the mostly radioactive actinides thorium (half-life of $\approx$14\,Gyr for $^{232}$Th) and uranium (half-life of $\approx$4.5\,Gyr for $^{238}$U) that are formed purely by $r$-process events, claims of the presence of these elements in the atmosphere of PS date back to as early as \citet{cowley1977}. Of the shockingly few ($\sim$200) Ap stars for which chemical abundances have been reported in the literature \citep{2018MNRAS.480.2953G}, PS leads the pack in terms of [Th/H] \citep[enhanced by between 2.75--3.89 dex relative to solar;][]{cowley2000,shulyak2010}. Similarly, the claimed [U/H] enhancement \citep[between 2.82--4.12 over solar;][]{cowley2000,shulyak2010} is topped only by the upper limit of [U/H]$<$4.88 that was reported without much ado for HD\,26385 \citep{bolcal1991}. The solar Th abundance itself is well constrained by meteoric estimates, but attempts to measure it from the Sun's photosphere rely on just one line \cite[\ion{Th}{ii}~4019\,{\AA}; e.g.,][]{2008A&A...483..591C} that is blended with several other far stronger lines.

The relatively warm \citep[$T_{\rm eff}=11\,000$\,K;][]{nielsen2020} and super-slowly rotating \citep[$P_{\rm rot}\approx21.8$ years;][]{1988A&A...199..299R} Ap star HR\,465 has often been discussed in the context of PS due to similarities such as the strong enhancement of heavy REE (Tb, Dy, Ho, Er) that are absent from the majority of Ap star spectra \citep[e.g.,][]{aikman1979,cowley1987}. In fact, HR\,465 represents the first literature mentions of Th and U detections in an Ap star spectrum. The claim of a U overabundance came first, with \citet{cowley1972} reporting that the strengths of three observed features attributed to U~{\sc ii} indicated a surface abundance roughly one million times that of the Sun. The Th overabundance claim came subsequently when \citet{cowley1975} attributed a feature at 4019\,{\AA} in the spectrum of HR\,465 to Th~{\sc ii}, with the line strength indicating a large overabundance, similar to U. 

A more recent study by \citet{nielsen2020} was unable to confirm the presence of Th and U in a modern spectrum of HR\,465 taken at a rotation phase of $\phi_{\rm rot}=0.68$. The Th~{\sc ii} lines seemed to be absent altogether, and while the wavelengths of two unknown features closely corresponded to positions of U~{\sc ii} lines, the authors stated that ``other stronger transitions from lower energy states do not support the presence of uranium in the stellar photosphere.'' Considering the highly variable abundances of HR~465 as a function of rotation phase, including a distinct ``REE maximum'' phase \citep{1988A&A...199..299R} that occurs around $\phi_{\rm rot}=0$, it is possible that Th and U were simply weak in the $\phi_{\rm rot}=0.68$ spectrum of \citet{nielsen2020}.

Searches for PS and HR\,465 analogues \citep[e.g.][]{hubrig2002} have turned up a few additional stars -- HD\,217522 and HD\,965 -- that share the peculiarity of an overwhelming REE spectrum and that have effective temperatures closer to that of PS \citep[6750\,K for HD\,217522 and 7500\,K for HD\,965;][]{2018MNRAS.480.2953G}. HD\,965 in particular has been involved in the recent PS lore due to claims of detection of promethium \citep[e.g.,][]{aller1970,cowley2004,fivet2007}, which is the only REE without any long-lived isotopes and hence the only REE that should not be present in a stellar atmosphere. Similar to PS, HD\,965 is a ssrAp star with a recently determined 16.5-year rotation period \citep{mathys2019} and with a relatively modest magnetic field strength of $\sim$4.3\,kG \citep{mathys1997,giarrusso2022}.

Two common themes among the reports of Th and U detections in Ap star spectra have been a reliance on wavelength coincidence statistics based on unresolved magnetically split lines and a lack of visual evidence provided (e.g., a plot showing numerous Th and U lines at the expected wavelengths and relative strengths). The best evidence to date of Th and U in stellar atmospheres has actually been provided by a handful of extremely metal-poor, $r$-process-enhanced Galactic halo stars, the most studied of which is probably CS\,31082-001 \citep[][]{2001Natur.409..691C,hill2002}. The presence of Th has been shown to be fairly unambiguous for this star, with several Th~{\sc ii} lines detected in blue/UV spectra, but the U detection relies on careful treatment of a small dent in the wing of a strong Fe~{\sc i} line that can be attributed to U~{\sc ii}. The Th and U abundances of these stars have been used to estimate stellar ages based on comparison to models of the Th and U production ratios in $r$-process events \citep[e.g.][]{2002ApJ...579..626S}.

Here, we use a combination of relatively high-resolution near-infrared (NIR) $H$-band spectra from the Apache Point Observatory Galactic Evolution Experiment \citep[APOGEE;][]{2017AJ....154...94M} and high-resolution optical spectra from the 8.2m Very Large Telescope (VLT) and its UV-Visual Echelle Spectrograph \citep[UVES;][]{2000SPIE.4008..534D} to establish by far the most convincing detection to date of thorium, and of an actinide in general, in an Ap star atmosphere. The presence of Th~{\sc iii} at high abundance on the surface of the Ap star {\cpd} is supported in both the NIR and optical data not only by matches between observed local minima and known Th~{\sc iii} wavelengths but also by matches between the observed Zeeman splitting patterns and those predicted from the atomic data. This Zeeman coincidence statistics (ZCS) method was first suggested as a line identification tool by \citet{1995ASPC...81..531M}, and it has been used previously to confirm the detection of neutral lithium in the spectra of cool Ap stars \citep{kochukhov2008} as well as to vet the Nd~{\sc iii} linelist \citep{ryabchikova2006}.

Little was known about {\cpd} prior to the discovery of resolved, magnetically split lines (RMSL) in $H$-band spectra \citep{choj2019} that indicated a mean magnetic field modulus around {\bmag}$\sim$8\,kG. The B8 temperature class reported by \citet{1976A&AS...23..283L} seems to be an error given that the 3rd data release of the Gaia mission \citep[DR3;][]{gaia,gaiadr3} indicates a far cooler $T_{\rm eff}=6834$\,K. Gaia DR3 also quotes a distance of 473\,pc, a logarithmic surface gravity of $\log g=3.92$, a stellar mass of 1.64\,$M_{\odot}$, and a stellar radius of 2.41\,$R_{\odot}$, all of which put {\cpd} in the parameter ballpark of PS (albeit with {\cpd} being more distant and thus fainter at $V\sim10.4$). {\cpd} has also been noted as a photometric variable, with for example the All-Sky Automated Survey for Supernovae \citep[ASAS-SN;][]{2018MNRAS.477.3145J} quoting a 518\,d period with an amplitude of 0.2\,mag. As will be seen, the rotation period implied by the multi-epoch spectra is considerably longer. 

The paper is laid out as follows. Section~\ref{data} provides a summary of the data used, while the associated analyses (including the use of time series mean magnetic field modulus measurements for constraining the rotation period of {\cpd}) are detailed in Section~\ref{analysis}. Sections~\ref{hband} and \ref{sec:hbandth3} focus on the $H$-band spectra, providing line identifications as well as evidence of $H$-band Th~{\sc iii} lines. Sections~\ref{opticalth} and \ref{opticalions} are similar, but focusing on the optical spectra and solidifying without a doubt the Th~{\sc iii} detection. The paper concludes in Section~\ref{compareth} with a discussion of additional stars whose archival UVES spectra show some evidence of Th~{\sc iii}. 

\begin{figure*}
\includegraphics[width=\textwidth]{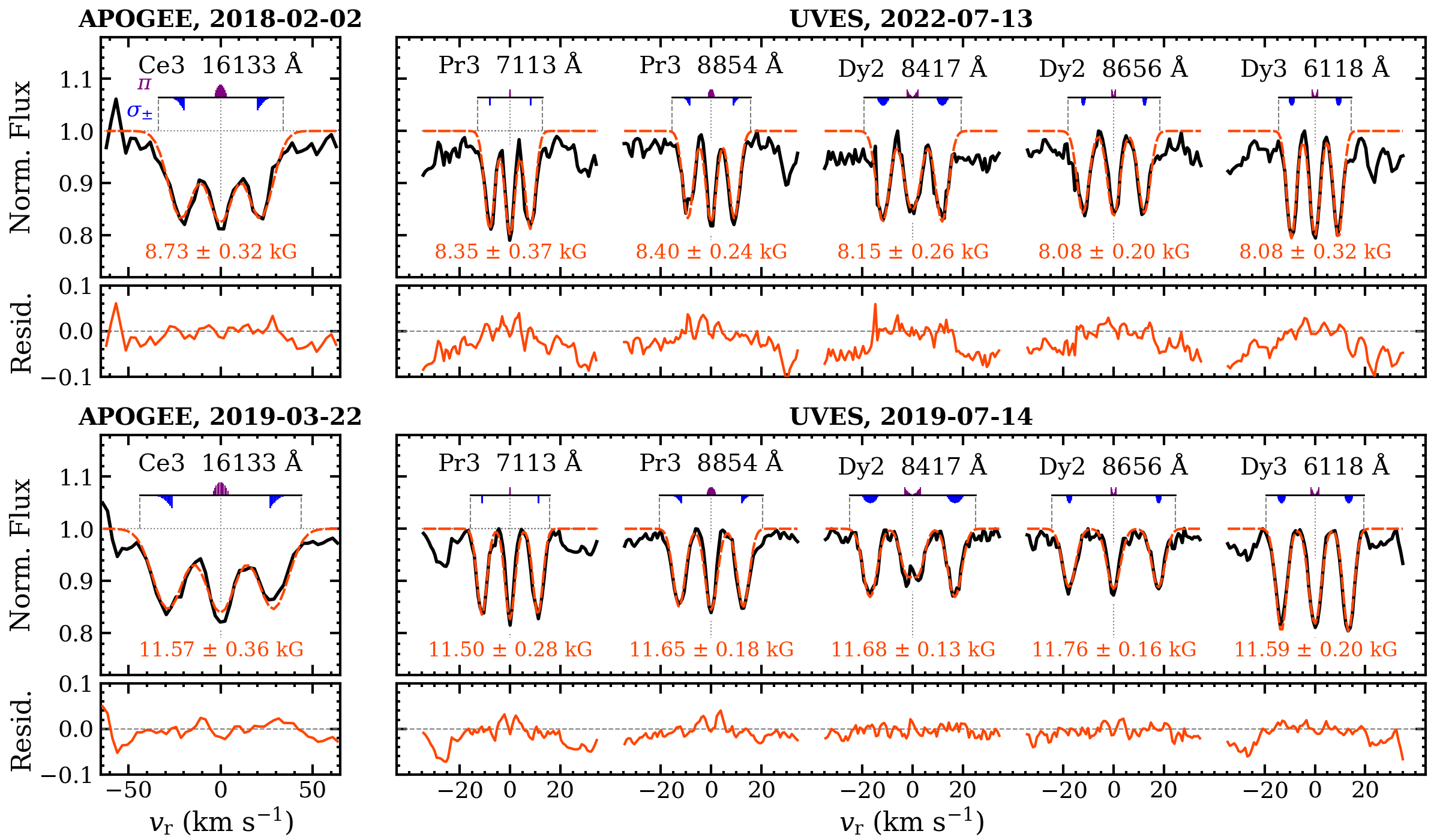}
\caption{Examples of Gaussian fits to the Ce~{\sc iii} 16133\,{\AA} line in two APOGEE spectra and to Pr~{\sc iii}, Dy~{\sc ii}, and Dy~{\sc iii} lines in two UVES spectra. Observed minus fit residuals are shown in the smaller panels. In the larger line profile panels, the individual $\pi$ (upward ticks) and $\sigma_{\pm}$ (downward ticks) Zeeman components are displayed above each line profile with the fixed relative intensities scaled arbitrarily and with the fixed relative velocity separations scaled to the {\bmag} obtained from the fit and quoted below each line profile. Vertical and horizontal dotted lines indicate line centers and continuum level, and vertical dashed lines demonstrate how the windows for direct summation equivalent widths are established on a line-by-line and {\bmag}-by-{\bmag} basis. \label{fig:zfit}}
\end{figure*}

\section{Spectroscopic Data} \label{data}
\subsection{APOGEE {\emph H}-band Spectroscopy} \label{apogee}
APOGEE is a component of the Sloan Digital Sky Survey \cite[SDSS; e.g.,][]{2020ApJS..249....3A} that has been operating on the Sloan 2.5-m telescope \citep{2006AJ....131.2332G} at Apache Point Observatory (APO, APOGEE-N) since 2011, and on the Ir\'{e}n\'{e}e du Pont 2.5-m telescope at Las Campanas Observatory (LCO, APOGEE-S) since 2017. The APOGEE instruments are duplicate 300-fiber, $R\approx22\,500$ spectrographs \citep{2019PASP..131e5001W} that record most of the $H$-band (15145--16960 {\AA}; vacuum wavelengths used throughout this paper when referring to the $H$-band) onto three detectors, with gaps between 15800--15860 {\AA} and 16430--16480 {\AA} due to non-overlapping wavelength coverage of the detectors. For a details of the APOGEE data reduction pipeline, see \citet{2015AJ....150..173N}, and for details of the APOGEE targeting strategy, see \citet{zasowski2013}. 

{\cpd} was observed by the APOGEE-S instrument a total of 23 times between 2018 February 2 and 2019 March 22, with six distinct groupings of observations separated by one or several days. A 24$^{\rm th}$ observation took place on 2023 April 27, just hours before we received the referee report on this paper. We decided to include this spectrum in the revised manuscript since the associated magnetic field strength estimate helps to constrain our estimate of the rotation period (see Section~\ref{prot}).

\subsection{VLT/UVES Optical Spectroscopy}
{\cpd} was observed by the UV-visual echelle spectrograph \citep[UVES;][]{2000SPIE.4008..534D} on the 8.2-m Kueyen unit of the Very Large Telescope (VLT) at the Cerro Paranal Observatory a total of eight times between 2019 April 4 and 2022 August 3. UVES uses a beam splitter to record the most of the optical wavelength range onto two CCDs, achieving resolutions of $R\approx74\,000$ in the blue when using a 0$\farcs$4 slit and $R\approx107\,000$ in the red when using a 0$\farcs$3 slit. We used a setup that resulted in the blue detector covering 3755--4980\,{\AA} and the red detector covering 5688--9459\,{\AA}. The data were reduced using version 6.1.3 of the European Southern Observatory (ESO) Reflex software. The UVES observations of {\cpd} are summarized in Table~\ref{tab:obs}. Due to extreme blending in the UVES blue coverage for {\cpd}, our analyses focus on the red coverage where contributions from iron peak elements and singly-ionized REE (REE2 from here on) are far fewer.

\begin{figure*}
\includegraphics[width=\textwidth]{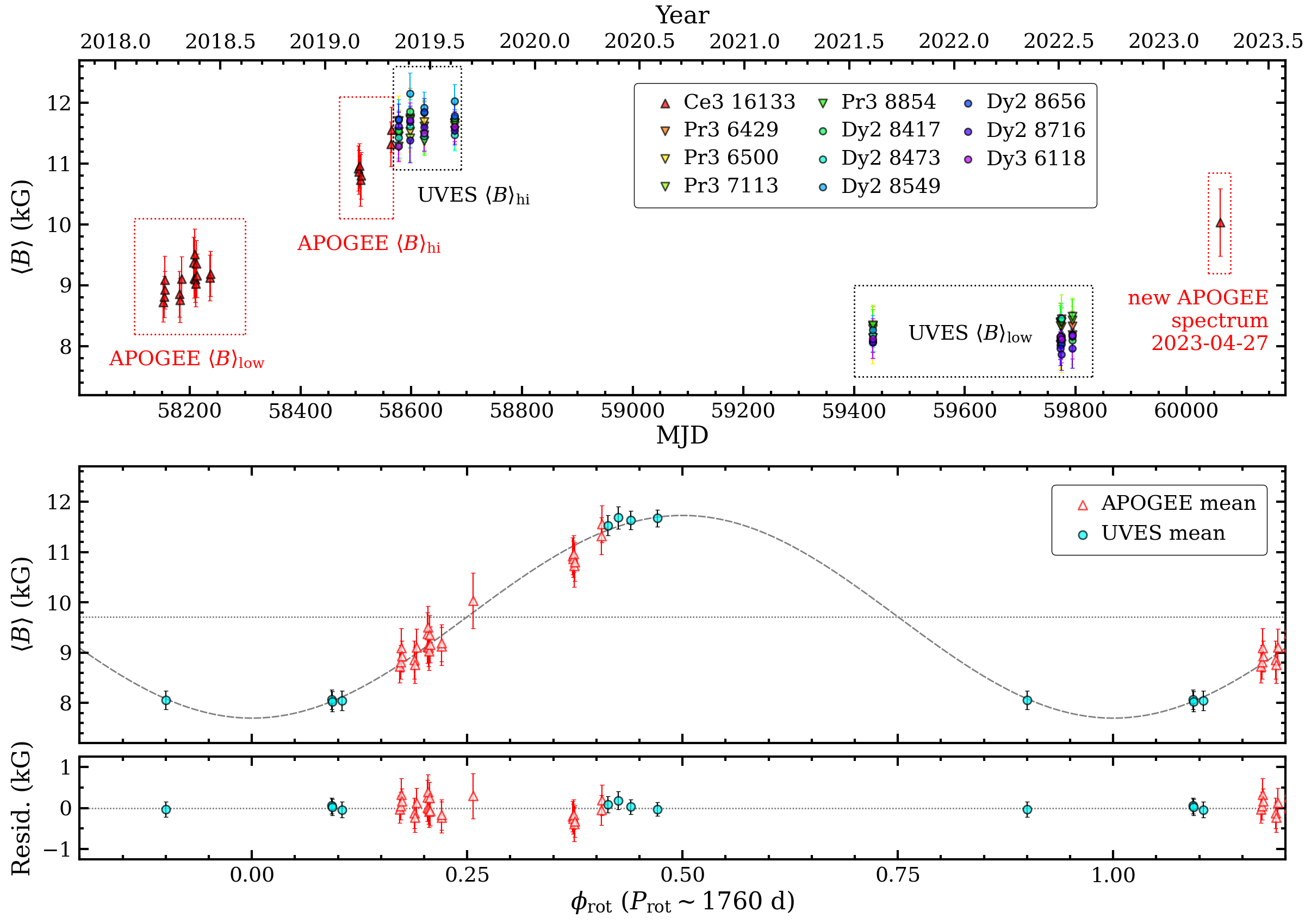}
\caption{\emph{Upper panel:} the line-by-line {\bmag} measurements of {\cpd} as a function of time. With the exception of the recent APOGEE observation at far right (excluded due to relatively low S/N and intermediate {\bmag}), the dotted line boxes demonstrate epoch groupings from which the associated spectra were combined. These $\langle B \rangle_{\rm low}$ and $\langle B \rangle_{\rm hi}$ spectra will be used for the remainder of this paper. \emph{Lower panels:} the epoch-averaged {\bmag} measurements of {\cpd} phased by a possible 1760 day rotation period, with the dashed curve in the larger panel representing the sine curve that best fits the measurements. Observed minus fit residuals are shown in the smaller panel. \label{fig:prot}}
\end{figure*}

\begin{table*}
\caption{Summary of spectroscopic observations of {\cpd}. The approximate rotation phases ($\phi_{\rm rot}$) were calculated based on the measurements of the mean magnetic field modulus ({\bmag}). Under the assumption of sinusoidal {\bmag} variation, the measurements can be satisfactorily fit using a rotation period of $P_{\rm rot}=$1760.4\,days and a modified julian data of MJD$=$57849.9 as an epoch of minimum magnetic field strength. \label{tab:obs}}
\begin{tabular}{cccrrrrr}
\hline
Date & MJD-Mid & Instrument & $t_{\rm exp}$ (s) & S/N & $v_{\rm r}$ (km s$^{-1}$) & $\langle B \rangle$ (kG) & $\phi_{\rm rot}$ \\ 
\hline
2018-02-02 & 58151.394 & APOGEE-S & 501 & \phantom{1}85 & $-1.84\pm0.83$ & $8.73\pm0.33$ & 0.17 \\ 
2018-02-04 & 58153.380 & APOGEE-S & 2002 & 177 & $-2.02\pm0.83$ & $8.81\pm0.34$ & 0.17 \\ 
2018-02-05 & 58154.336 & APOGEE-S & 1502 & 103 & $-1.93\pm1.05$ & $9.09\pm0.40$ & 0.17 \\ 
2018-02-06 & 58155.289 & APOGEE-S & 1001 & \phantom{1}90 & $-2.29\pm0.76$ & $8.93\pm0.31$ & 0.17 \\ 
2018-03-04 & 58181.284 & APOGEE-S & 2002 & 165 & $-2.19\pm0.98$ & $8.86\pm0.38$ & 0.19 \\ 
2018-03-05 & 58182.289 & APOGEE-S & 1502 & 160 & $-1.74\pm0.94$ & $8.76\pm0.36$ & 0.19 \\ 
2018-03-08 & 58185.235 & APOGEE-S & 2503 & 202 & $-1.86\pm0.93$ & $9.11\pm0.36$ & 0.19 \\ 
2018-03-30 & 58207.179 & APOGEE-S & 2002 & 167 & $-2.53\pm1.18$ & $9.38\pm0.42$ & 0.20 \\ 
2018-03-31 & 58208.193 & APOGEE-S & 2002 & 174 & $-1.31\pm0.81$ & $9.12\pm0.32$ & 0.20 \\ 
2018-04-01 & 58209.124 & APOGEE-S & 2002 & 180 & $-2.44\pm1.21$ & $9.51\pm0.42$ & 0.20 \\ 
2018-04-02 & 58210.105 & APOGEE-S & 2002 & 148 & $-2.16\pm1.00$ & $9.10\pm0.38$ & 0.21 \\ 
2018-04-03 & 58211.165 & APOGEE-S & 2503 & 169 & $-1.94\pm1.00$ & $9.03\pm0.38$ & 0.21 \\ 
2018-04-04 & 58212.110 & APOGEE-S & 2002 & 170 & $-2.00\pm1.02$ & $9.36\pm0.38$ & 0.21 \\ 
2018-04-05 & 58213.170 & APOGEE-S & 2002 & 170 & $-1.85\pm0.94$ & $9.16\pm0.36$ & 0.21 \\ 
2018-04-28 & 58236.094 & APOGEE-S & 1502 & 123 & $-1.98\pm0.96$ & $9.13\pm0.38$ & 0.22 \\ 
2018-04-29 & 58237.108 & APOGEE-S & 1502 & 102 & $-1.45\pm0.94$ & $9.19\pm0.37$ & 0.22 \\ 
2019-01-21 & 58504.318 & APOGEE-S & 2503 & 166 & $-1.71\pm0.87$ & $10.92\pm0.37$ & 0.37 \\ 
2019-01-22 & 58505.332 & APOGEE-S & 3003 & 200 & $-1.78\pm0.84$ & $10.86\pm0.36$ & 0.37 \\ 
2019-01-23 & 58506.369 & APOGEE-S & 2002 & 141 & $-1.66\pm0.88$ & $10.96\pm0.37$ & 0.37 \\ 
2019-01-25 & 58508.291 & APOGEE-S & 2002 & 125 & $-1.95\pm1.03$ & $10.73\pm0.42$ & 0.37 \\ 
2019-01-26 & 58509.297 & APOGEE-S & 2503 & 137 & $-1.67\pm0.89$ & $10.80\pm0.38$ & 0.38 \\ 
2019-03-21 & 58563.120 & APOGEE-S & 1502 & 132 & $-2.00\pm0.84$ & $11.32\pm0.37$ & 0.41 \\ 
2019-03-22 & 58564.127 & APOGEE-S & 2002 & 160 & $-1.71\pm0.86$ & $11.56\pm0.37$ & 0.41 \\ 
2023-04-27 & 60061.112 & APOGEE-S & 447 & \phantom{1}58 & $-1.86\pm1.42$ & $10.04\pm0.55$ & 0.26 \\ 
2019-04-04 & 58577.137 & UVES & 1200 & \phantom{1}72 & $-1.56\pm0.10$ & $11.53\pm0.20$ & 0.41 \\ 
2019-04-25 & 58598.051 & UVES & 1200 & \phantom{1}59 & $-1.77\pm0.11$ & $11.68\pm0.22$ & 0.43 \\ 
2019-05-20 & 58623.018 & UVES & 1200 & 106 & $-1.22\pm0.09$ & $11.63\pm0.18$ & 0.44 \\ 
2019-07-14 & 58678.053 & UVES & 1200 & 124 & $-1.36\pm0.08$ & $11.67\pm0.16$ & 0.47 \\ 
2021-08-07 & 59433.991 & UVES & 1400 & \phantom{1}89 & $-1.53\pm0.09$ & $8.19\pm0.18$ & 0.90 \\ 
2022-07-12 & 59772.006 & UVES & 1680 & \phantom{1}63 & $-1.46\pm0.10$ & $8.17\pm0.20$ & 0.09 \\ 
2022-07-13 & 59773.994 & UVES & 1680 & \phantom{1}72 & $-2.08\pm0.10$ & $8.20\pm0.20$ & 0.09 \\ 
2022-08-03 & 59794.010 & UVES & 1680 & \phantom{1}92 & $-1.61\pm0.10$ & $8.22\pm0.20$ & 0.10 \\ 
\hline
\end{tabular}
\end{table*}

\subsection{Linelist}
The optical and $H$-band rest wavelengths, oscillator strengths ($\log gf$), and effective Land\'{e} factors ($g_{\rm eff}$) used here were taken primarily from the Vienna Atomic Line Database \citep[VALD;][]{2019ARep...63.1010P}. Although VALD contains lines for most of the doubly-ionized REE (REE3 from here on), ions like Sm~{\sc iii} and Gd~{\sc iii} are glaring exceptions that likely account for a non-negligible fraction of the plethora of unidentified lines produced by {\cpd}. Another exception is Ho~{\sc ii}, which is strong for {\cpd} and for which the National Institute of Standards and Technology \citep[NIST;][]{NIST_ASD} atomic database lists far more lines (albeit lacking most of the important atomic data) than does VALD. U~{\sc iii} is another unfortunate omission, since it may also be responsible for some fraction of the unidentified lines and since its detection might allow for a radioactive cosmochronological age estimate for {\cpd}. 

Given that this paper presents the first unambiguously demonstrated proof of Th~{\sc iii} lines in a stellar spectrum, it does not suffice to simply credit VALD. The optical portion of the Th~{\sc iii} linelist was determined experimentally by \citet{2002ApJ...567.1276B}, and these authors correctly hypothesized that some Th~{\sc iii} lines should appear in the spectra of Ap stars with large overabundances of REE3. The thorium linelist was later extended into the infrared by \citet{2014ApJS..211....4R}, with seven of of the Th~{\sc iii} transitions falling in APOGEE's coverage of the $H$-band.

\section{Analysis} \label{analysis}
\subsection{Magnetic Field Measurement}
The spectra of {\cpd} were analyzed with a Python code (to be made publicly available pending some improvements and full documentation) that performs least-squares fitting of the observed line profiles of one or more atomic transitions with radial velocity ($v_{\rm r}$), mean magnetic field modulus ({\bmag}), full width at half maximum (FWHM), and equivalent width ($W_{\lambda}$) as the variables to be either fixed or allowed to vary. The individual $\pi$ and $\sigma_{\pm}$ Zeeman components of each transition, as determined based on the total angular momentum quantum numbers ($J$; which dictate the number of components and their relative strengths) and Land\'{e} factors ($g$; which dictate the magnitude of the splitting) of the upper and lower energy levels, are approximated as Gaussians of uniform FWHM. The relative strengths and separations of the Zeeman components are fixed to those predicted by the atomic data but scaled in wavelength or velocity space in order to match the observed line profile and thus estimate {\bmag}. 

Once a best fit has been achieved, {\bmag} is calculated via {\bmag}$=\Delta\lambda \, \lambda_{0}^{-2} \, k^{-1} \, g_{\rm eff}^{-1}$, where $\lambda_{0}$ is the rest wavelength of the transition, $\Delta\lambda$ are the separations of the Zeeman components from the line center, $k$ is a constant equal to $4.671\times10^{-13}$\,{\AA}$^{-1}$\,G$^{-1}$, and $g_{\rm eff}$ is the effective Land\'{e} factor. This method is an approximation for all but the simplest cases, but it nonetheless provides {\bmag} estimates that are roughly consistent for a variety of Zeeman patterns. 

The analysis focused on just one line in the $H$-band -- Ce~{\sc iii} 16133.17\,{\AA}. In addition to being one of the strongest, most isolated, and most clearly split lines covered, the Ce~{\sc iii} strength is relatively constant from epoch to epoch unlike some other ions. Several other lines meet the same criteria, but unfortunately their carrier ion(s) is unidentified, with no likely counterparts in the available atomic data.

Fe~{\sc ii}~6149.246\,{\AA} is widely regarded as the ground truth for {\bmag} measurement \citep{mathys2017}, and indeed is the most likely RMSL to be present in Ap/Bp spectra due to having $J=0.5$ for both levels of the transition, $g_{\rm lo}=0$, and $g_{\rm hi}=2.7$. This combination of $J$ and $g$ results in a simple Zeeman doublet pattern that is highly sensitive to field strength due to $g_{\rm hi}$ taking the place of $g_{\rm eff}$ in the aforementioned equation. Unfortunately, the line is problematic in this case. First of all, the blending is quite severe, with possible contributions from an Sm~{\sc ii} line at 6149.060\,{\AA} and with definite contributions from an unidentified line at 6148.85\,{\AA}. Although satisfactory fits to the Fe~{\sc ii}$+$Sm~{\sc ii}$+$unknown blend can often be achieved for stars with weaker magnetic fields by treating the unknown line as a single wide Gaussian \citep[e.g.][]{giarrusso2022}, that approximation is in this case complicated by the strong and variable surface magnetic field strength, the unknown Zeeman pattern of the 6148.85\,{\AA} line, the variable line strengths of most ions, and the fact that some ions exhibit significant phase lags in the variability with respect to others.  

Instead, {\bmag} was estimated from the UVES spectra using ten isolated REE lines that were present during all epochs and widely split into quasi-triplet features. This included four Pr~{\sc iii} lines (6429, 6500, 7113, 8854\,{\AA}), five Dy~{\sc ii} lines (8417, 8473, 8549, 8656, 8716\,{\AA}), and one Dy~{\sc iii} line (6118.4\,{\AA}). Figure~\ref{fig:zfit} shows examples of line profile fitting in the two APOGEE spectra and two UVES spectra that represent the observed {\bmag} extremes from each data set. 

\subsection{Rotation Period} \label{prot}
The {\bmag} measurements from the APOGEE spectra show that the field strength of {\cpd} increased steadily from around $\sim$8.7\,kG in February of 2018 up to $\sim$11.5\,kG in late March of 2019. This trend continued in the UVES spectra from April, May, and July of 2019, with {\bmag} hovering around 11.6\,kG. More than two years passed before the next UVES observation in August of 2021, and by then {\bmag} reached a new low of 8.2\,kG. The three spectra from July--August of 2022 were all but identical to the 2021 spectrum, with {\bmag} still hovering around the observed minimum of 8.2\,kG. The most recent APOGEE observation from late April of 2023 indicated that {\bmag} had climbed back up to $\sim$10.0\,kG once again.

\begin{figure*}
\includegraphics[width=\textwidth]{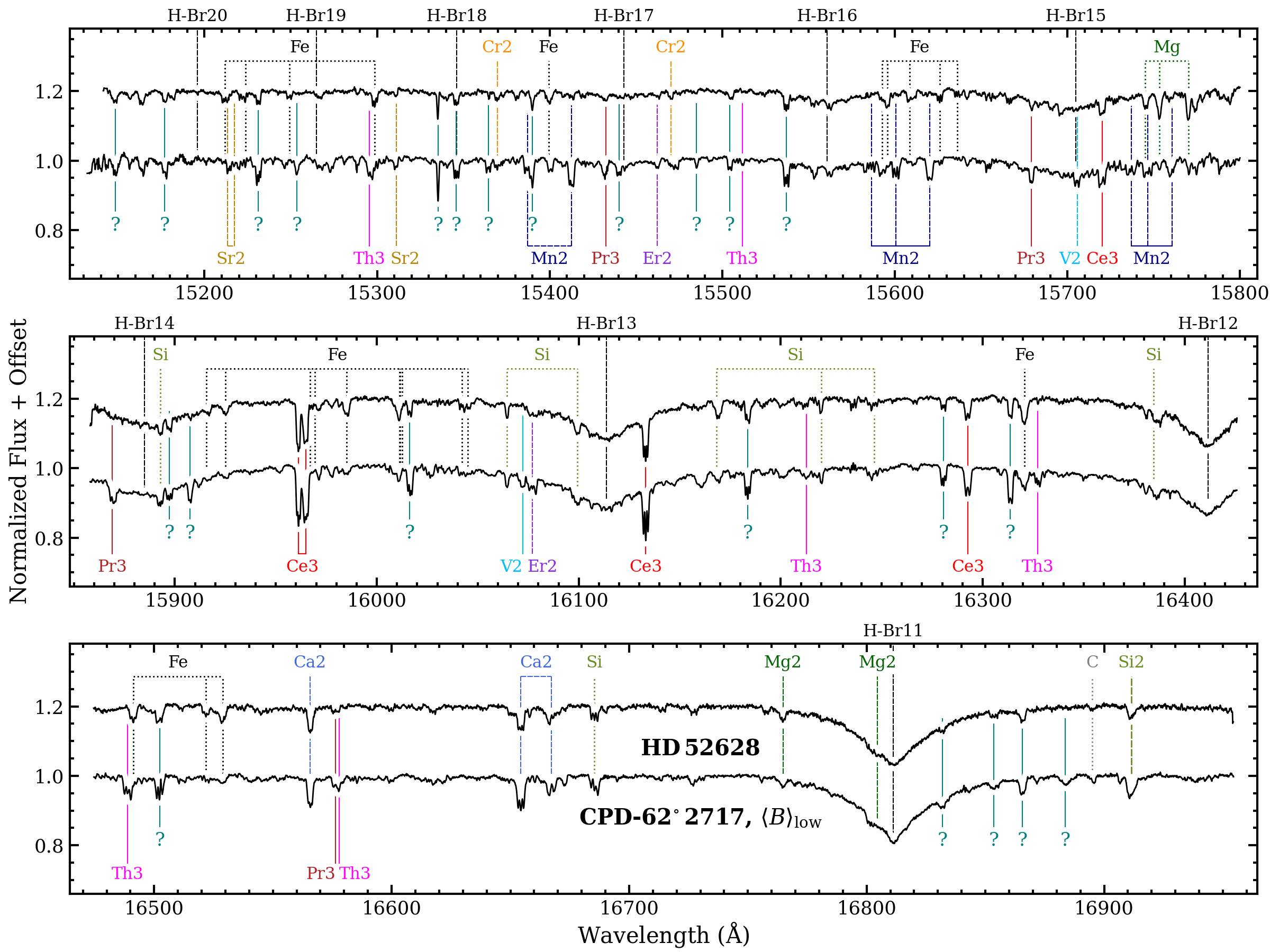}
\caption{The mean APOGEE {\blo} spectrum of {\cpd} compared to the mean APOGEE spectrum of HD\,52628. Most of the strong absorption lines are labeled, with ``H-Br11--20'' indicating the hydrogen Brackett series lines and with question marks indicating lines that lack likely counterparts in the existing atomic data. For conciseness, ionization stages are omitted from neutral elements labels and given as integers for ionized elements.\label{fig:hspec}}
\end{figure*}

All of the above indicates an exceedingly long rotation period for {\cpd}, and this is demonstrated in Figure~\ref{fig:prot}. The upper panel shows the {\bmag} measurements as a function of time, with dotted line boxes enclosing epoch groupings in which the combined spectra will be used for the remainder of this paper. The 2023 April APOGEE spectrum is not included in these groupings since it has considerably lower S/N than the other APOGEE spectra and since the associated intermediate {\bmag} it does not fit coherently with either the {\bmag}$_{\rm low}$ nor {\bmag}$_{\rm hi}$ epoch groupings.

The best-fitting sine curve to the {\bmag} measurements is shown in the lower panel of Figure~\ref{fig:prot}, and it is admittedly tentative due to the sparseness of the observations and the large gaps in phase coverage. Nonetheless, the available data indicate $P_{\rm rot}\approx1760$\,days (4.82 years), which places {\cpd} firmly into the class of super-slow-rotating Ap stars (ssrAp; Ap stars with $P_{\rm rot}>50$ days) that was recently defined by \citet{2020pase.conf...35M}. Considering that we found $P_{\rm rot}\approx$1765\,days prior to the APOGEE spectrum from 2023 April, addition of that data point merely validated rather than changed the results.

\subsection{Radial Velocities}
As for the radial velocities, we found an average from the APOGEE spectra of $v_{r}=-1.91\pm0.99$\,km\,s$^{-1}$ (after applying heliocentric corrections) and a far more precise average of $v_{r}=-1.57\pm0.10$\,km\,s$^{-1}$ from the UVES spectra, for an overall average of $v_{r}=-1.83\pm0.83$\,km\,s$^{-1}$. The individual epoch measurements range from $-2.53$ to $-1.22$\,km\,s$^{-1}$, but they all agree to within the error bars such that we have no reason to suspect {\cpd} having a binary companion. If it does, either the mass ratio must be very low or the orbital period very long. 

Table~\ref{tab:obs} provides a summary of the observations of {\cpd}, including observation dates, Modified Julian Dates (MJD) at mid-exposure, instruments used, exposure times, signal-to-noise ratios (S/N; as reported by the APOGEE and UVES pipelines), heliocentric radial velocities ($v{\rm r}$; in km\,s$^{-1}$), {\bmag} estimates, and approximate rotation phases ($\phi_{\rm rot}$).

\subsection{TESS Lightcurve} 
Considering the remarkably high fraction of roAp stars found among the ssrAp stars (22\%, versus 3\% for non-ssrAp stars) by \citet{mathys2020b}, it is natural to wonder if {\cpd} might also be a short-period pulsator and whether this could be discovered from Transiting Exoplanet Survey Satellite \citep[TESS;][]{2015JATIS...1a4003R} data. To date, {\cpd} has been observed in three TESS sectors (1800\,s cadence in sector 11 in 2019 and 600\,s cadence in sectors 37 and 38 in 2021). 

Despite the cadences not being well-suited for detecting very short period variability, we used the \textsc{lightkurve} \citep{Lightkurve2018} package to download the associated lightcurves, all of which show perfectly repeating variability with an amplitude of a few percent and with a period of 0.5772 days as determined using the time series analysis code \textsc{Period04} \citep{Lenz2005}. This period is several dozen times longer than the longest known roAp pulsation period and can likely be attributed a nearby neighbor (22\arcsec separation; similar to the size of a TESS pixel) of comparable brightness (SIMBAD quotes $V=10.02$ for TYC 8979-1364-1 and $V=10.42$ for {\cpd}) that is almost certainly an RR Lyrae variable. Given the lack of short-cadence observations and contamination of the existing observations by the neighbor star, it is therefore difficult/impossible to determine whether or not {\cpd} is a roAp star based on the existing TESS data.

\section{$H$-band Spectra} \label{hband}
In the opinion of the first author of this paper, who has spent a great deal of time visually inspecting spectra of peculiar stars observed by the APOGEE instruments, {\cpd} is not particularly remarkable at first glance. Admittedly, the absorption lines are all magnetically split, thus placing {\cpd} in the select group of just a few hundred stars known to exhibit RMSL. However, the APOGEE Ap/Bp star sample contains a few dozen stars with {\bmag}$>8$\,kG and up to 24\,kG. Further, whereas many of the strongest observed lines in the {\cpd} spectra have no counterparts in the atomic data and instead remain unidentified, the same is true for several hundred other Ap/Bp stars in the APOGEE sample. 

\begin{figure}
\includegraphics[width=\columnwidth]{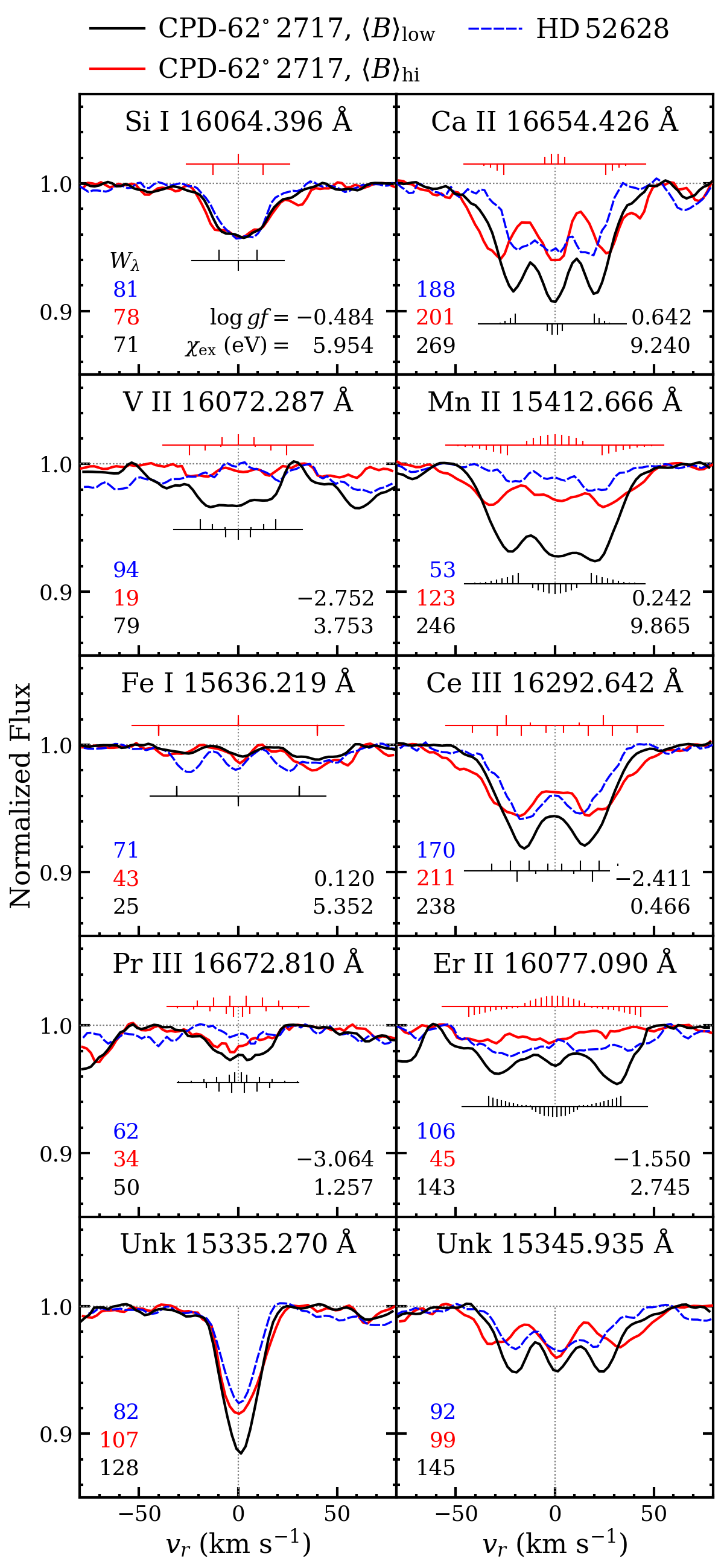}
\caption{Comparison of line profiles from the mean APOGEE {\blo} and {\bhi} spectra of {\cpd} and the mean APOGEE spectrum of HD\,52628. In cases where the lines could be identified, the individual $\pi$ and $\sigma_{\pm}$ components are shown above and below the line profiles, with the upper Zeeman pattern scaled horizontally to the {\bmag}$\sim$11.0\,kG of {\cpd} in {\bhi} mode, and with the lower, reflected Zeeman pattern scaled to the {\bmag}$\sim$8.8\,kG of {\cpd} in {\blo} mode. Direct summation equivalent widths ($W_{\lambda}$, in m{\AA}) are given below and left of each line profile, and oscillator strengths ($\log gf$) and excitation energies ($\chi_{\rm ex}$, in eV) are given below and right of the line profiles. \label{fig:hprofs}}
\end{figure}

These sentiments are demonstrated in Figure \ref{fig:hspec}, which displays the mean {\blo} APOGEE spectrum of {\cpd} along with the mean APOGEE spectrum of HD\,52628, which is the closest thing to a spectroscopic twin of {\cpd} that we could find. The similarity is striking in terms of both line content and line strengths, and the magnetic field strengths are also in the same ballpark ({\bmag}$\sim7$\,kG for HD\,52628). As is typical in the $H$-band spectra of Ap stars, the three strongest metal lines for both stars are Ce~{\sc iii} 15961, 15965, and 16133\,{\AA}. For {\cpd} and HD\,52628, the next strongest lines are Ca~{\sc ii} 16566 and 16654\,{\AA}, which are followed by numerous unidentified lines that we strongly suspect the majority of can be attributed to a single ion, most likely an REE3. Despite being unidentified, the unknown line at 15335\,{\AA} is particularly useful given that it clearly has an extremely low effective Land\'{e} factor. The line thus appears as a narrow spike in the case of slow rotation, and in addition to setting an upper limit on $v \sin i$, the line can be used to confirm magnetic splitting in other lines, all of which have considerably higher effective Land\'{e} factors. 

Some of the weaker lines that can actually be identified and that are in common between {\cpd} and HD\,52628 include a C~{\sc i} line, a few dozen Si~{\sc i} and Fe~{\sc i} lines, an Si~{\sc ii} line, a few Mg~{\sc ii} lines, several Pr~{\sc iii} lines, and a few weak Er~{\sc ii} lines. It was not until preparing this paper that we recognized the presence of the latter, and if not for the subject matter at hand, the Er~{\sc ii} lines at 15462\,{\AA} and 16077\,{\AA} would represent the heaviest metal ever detected in stellar $H$-band spectra (regardless of spectral type).

Despite the apparent similarity of {\cpd} and HD\,52628 when the spectra are viewed widely as in Figure \ref{fig:hspec}, it is the subtle differences that set {\cpd} apart from any other APOGEE-observed stars that we are aware of.

Figure~\ref{fig:hprofs} provides a detailed view of individual lines in the spectra displayed in Figure \ref{fig:hspec} as well as in the {\bhi} spectrum of {\cpd}. Si~{\sc i} 16064\,{\AA} is one of the few low-$g_{\rm eff}$ lines present in the {\cpd} spectra (along with the unknown 15335\,{\AA} line) for which magnetic splitting is not resolved at this combination of resolving power and magnetic field strength. Si~{\sc i} is also one of the few ions whose line strengths are relatively stable across the {\blo} and {\bhi} phases of {\cpd}. 

\begin{figure}
\includegraphics[width=\columnwidth]{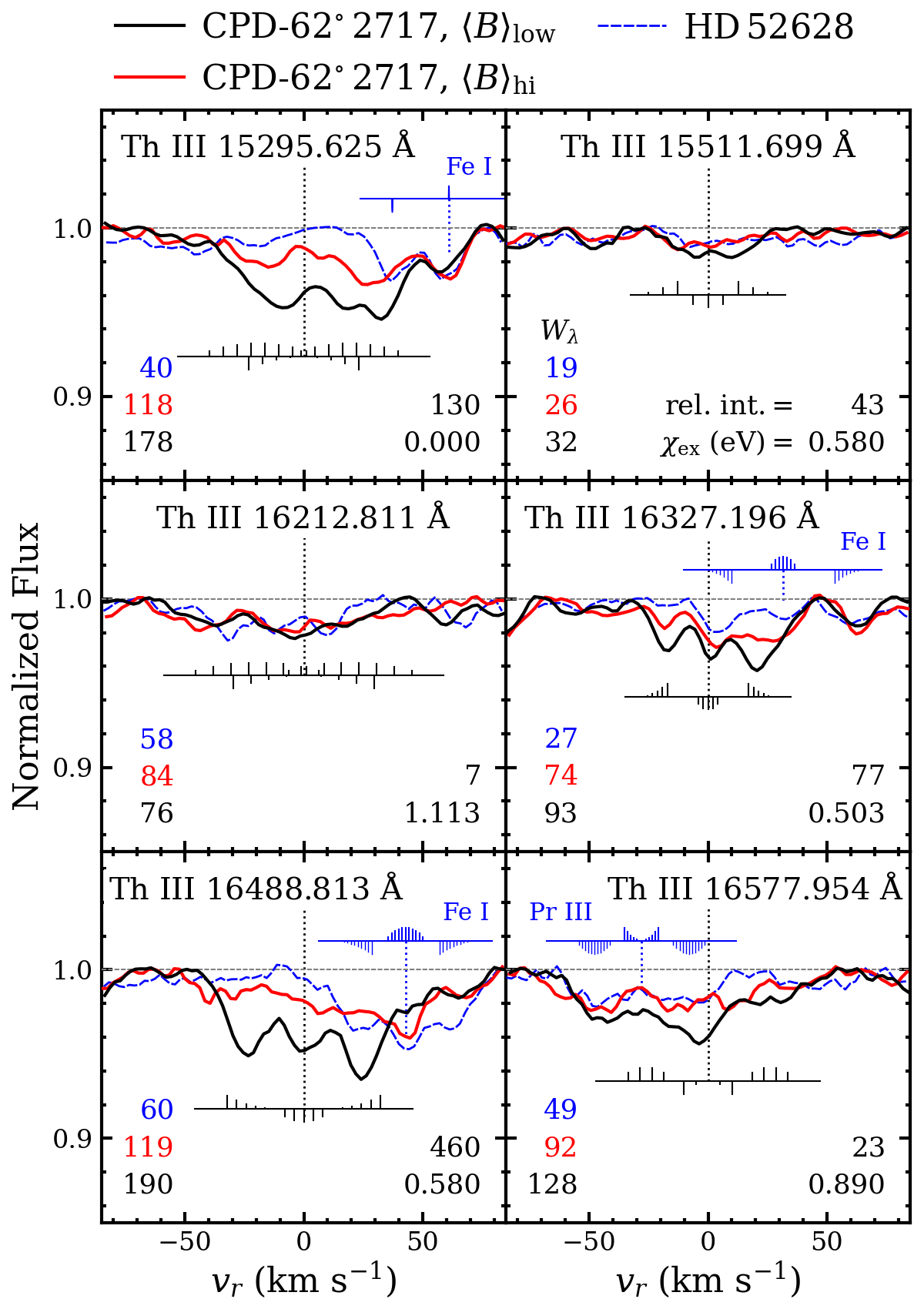}
\caption{The $H$-band Th~{\sc iii} lines of {\cpd}. Meanings are mostly the same as in Figure~\ref{fig:hprofs} except for a few details. Relative intensities from \citet{2014ApJS..211....4R} are given rather than $\log gf$, and to demonstrate that the Th~{\sc iii} lines are not actually present in the HD\,52628 spectrum, the Zeeman patterns of nearby Fe~{\sc i} and Pr~{\sc iii} lines are shown above the line profiles (scaled horizontally for the 7.0\,kG of HD\,52628). \label{fig:hprofsth3}}
\end{figure}

Beginning with Ca~{\sc ii}, the large degree of variability as a function of rotation phase of some of the other ions becomes apparent. The behavior of V~{\sc ii} and Mn~{\sc ii} is particularly noteworthy in this context due to the dramatic variability. The line strengths of both ions change by a factor of two or more between {\blo} and {\bhi} phases, and the V~{\sc ii} lines are essentially absent altogether from the {\bhi} spectrum. As for Fe~{\sc i}, the lines are quite weak in the {\cpd} spectra, but a hint can be seen for of variability that is anti-phased with respect to V~{\sc ii} and Mn~{\sc ii}. This is possibly true for Si~{\sc i} as well. 

The variability of the known and suspected REE is also a mixed bag, differing from ion to ion. Whereas Pr~{\sc iii}, Ce~{\sc iii}, and the unknown lines are relatively stable, Er~{\sc ii} follows the trend established by V~{\sc ii} and Mn~{\sc ii} in terms of the lines all but disappearing during {\bhi} phases.

\section{$H$-band Th~{\sc iii} Detection} \label{sec:hbandth3}
As remarkable as {\cpd} is without even mentioning thorium, Figure~\ref{fig:hprofsth3} demonstrates why the star is particularly special. Six of the seven Th~{\sc iii} lines from \citet{2014ApJS..211....4R} that are covered by APOGEE spectra are shown. The Th~{\sc iii} line that is not shown and that can be considered a non-detection falls at a vacuum rest wavelength of 16064.373\,{\AA}. In addition to \citet{2014ApJS..211....4R} having quoted a relative intensity of just 7 (a very low number in this context) for this line, it would be fully blended with Si~{\sc i}~16064.396\,{\AA} even if it was stronger. \citet{2014ApJS..211....4R} also quoted a relative intensity of 7 for the Th~{\sc iii} line at 16212.811\,{\AA}, and we consider it a tentative detection at best.

As for the five Th~{\sc iii} lines with relative intensities $>7$, they all appear to be present despite some level of blending in most cases. The \citet{2014ApJS..211....4R} linelist suggests the 15511.699\,{\AA} and 16577.954\,{\AA} lines should be quite weak, and indeed they are. Likewise, the 15295.625\,{\AA}, 16327.196\,{\AA}, 16488.813\,{\AA} lines should be the strongest $H$-band Th~{\sc iii} features, and indeed they are blatantly present albeit with the red wing of the 15295.625\,{\AA} line being blended with one of the strongest Fe~{\sc i} lines covered by the APOGEE spectra. For the 16327.196\,{\AA} and 16488.813\,{\AA} lines in particular, the agreement between the expected and observed central positions, relative strengths, and Zeeman patterns during the {\blo} phases of {\cpd} leaves little doubt as to the Th~{\sc iii} identification. 

Similar to V~{\sc ii}, Mn~{\sc ii}, and Er~{\sc ii}, the contributions from Th~{\sc iii} are sufficiently weak in the {\bhi} spectra of {\cpd} that they likely would not have been acknowledged were it not for the {\blo} APOGEE observations. It could therefore be argued that {\cpd} is perhaps not as unique as this paper implies, and that the observation timing has simply been unfortunate for the other hundreds of highly peculiar Ap stars that APOGEE has observed. As it stands for now however, {\cpd} can just as easily be considered a needle in a haystack due to an obviously severe Th~{\sc iii} overabundance patch on its surface along with a magnetic field strength suitable for confirmation via Zeeman pattern matches.

\begin{figure*}
\includegraphics[width=\textwidth]{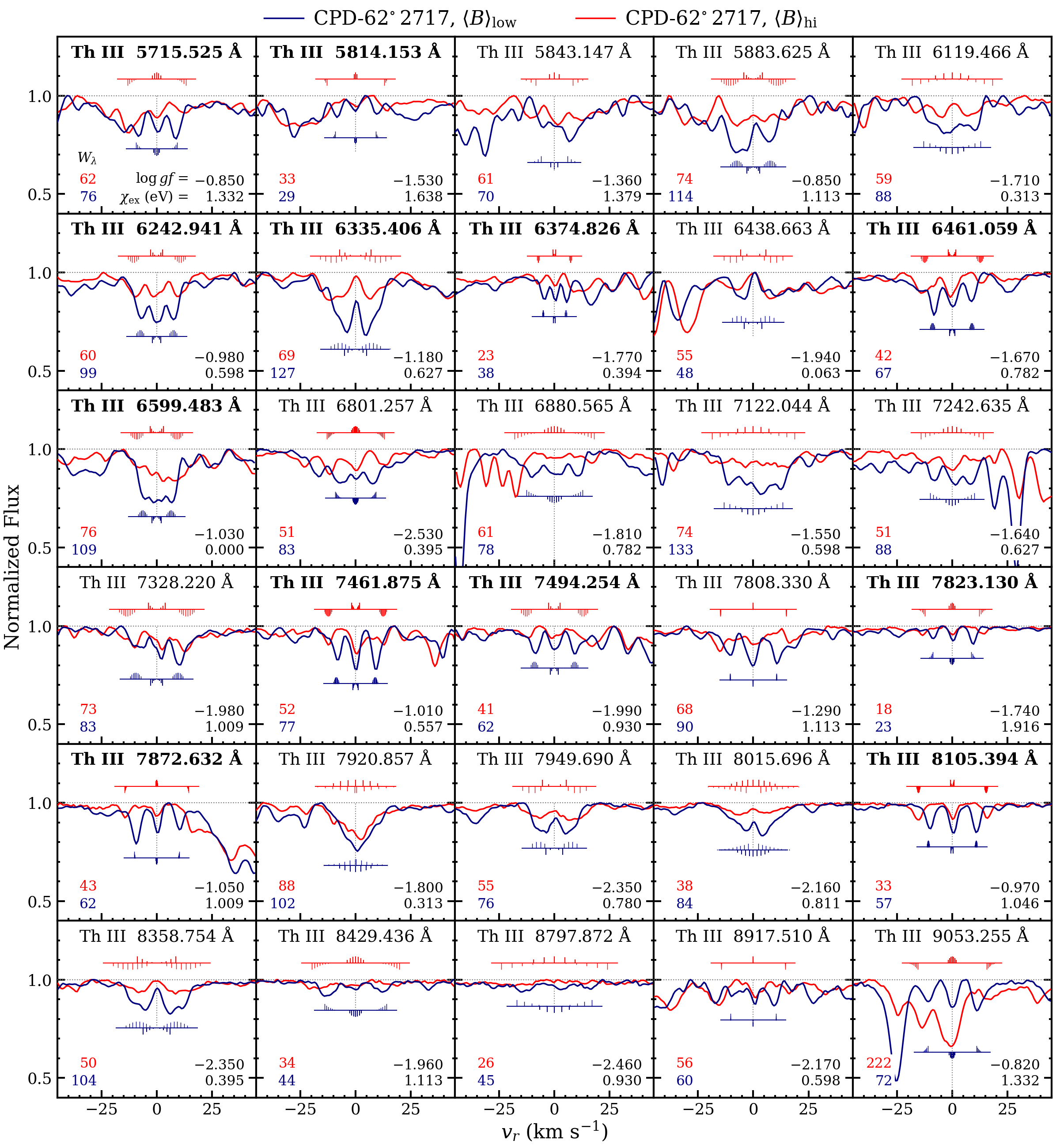}
\caption{The best evidence to date of thorium in the atmosphere of an Ap star. Each panel shows Th~{\sc iii} line profiles from the mean UVES {\blo} (navy) and {\bhi} (red) spectra of {\cpd}. The associated Zeeman patterns are shown above and below each line profile pair and scaled to {\bmag}$=11.6$\,kG (upper Zeeman pattern) and {\bmag}$=8.2$\,kG (reflected lower Zeeman pattern). Bold font of the labels above the line profiles indicates cases of minimally blended lines with relatively high $\log gf$ and for which the observed Zeeman patterns match those expected from the atomic data. As in Figure~\ref{fig:hprofs}, direct summation equivalent widths, oscillator strengths, and excitation energies are given below the line profiles.  \label{fig:oprofsth3}}
\end{figure*}

\section{Optical Th~{\sc iii} Detection} \label{opticalth}
Whereas the $H$-band spectra of {\cpd} merely provided hints about the degree of peculiarity at hand, the optical spectra make it clear that `extreme' is a proper qualifier. All of the trends hinted at by the $H$-band are confirmed, and we begin our discussion of the optical spectra by getting right to the point. Th~{\sc iii} lines are not only present in the UVES spectra of {\cpd}, they are numerous.  

The optical Th~{\sc iii} linelist provides atomic data for a total of 82 lines between 5700--9100\,{\AA}. Of these 82 lines, we see no evidence in the {\cpd} spectra of the 38 with excitation energies higher than 2\,eV and/or with oscillator strengths less than $-3.05$. Another 14 lines are either contaminated by telluric absorption during all observed epochs or else fall at roughly the same wavelengths of stronger lines from other ions. In Figure~\ref{fig:oprofsth3}, the line profiles of the remaining 30 Th~{\sc iii} lines are displayed as they appear in the UVES {\blo} and {\bhi} spectra of {\cpd}. Some evidence can be seen for the presence of all of them, despite most being blended to some extent.

Among the detected Th~{\sc iii} lines are a resonance line at 6599.483\,{\AA} (see Fig.~\ref{fig:oprofsth3}) that has an oscillator strength of $\log gf=-1.030$ and that should be among the strongest Th~{\sc iii} lines between 5700--9100\,{\AA}. Indeed it is one of them. The equivalent width of the 6599\,{\AA} line in the {\blo} is topped only by that of the Th~{\sc iii} lines at 6335\,{\AA} and 7122.044\,{\AA}, and the latter is almost certainly blended with an unknown line. The most likely contaminants of the 6599\,{\AA} line are the Fe~{\sc ii} lines at 6599.804\,{\AA} ($\log gf=-2.883$, $\chi_{\rm ex}=7.6973$\,eV) and 6600.025~{\AA} ($\log gf=+0.295$, $\chi_{\rm ex}=11.0164$\,eV), each of which could overlap the Th~{\sc iii} line sufficiently fast stellar rotation or sufficient magnetic splitting. In the case of {\cpd} however, the contributions to Th~{\sc iii}~6599\,{\AA} from Fe~{\sc ii} can be deduced to be minimal. Other Fe~{\sc ii} lines with longer wavelengths (and hence less chances of blending with REE lines), similar energy levels, and even higher $\log gf$ are either absent from the spectra or exceedingly weak. 

As noted above, the 6335.406\,{\AA} line (see Fig.~\ref{fig:oprofsth3}) is also among the strongest Th~{\sc iii} features in the {\cpd} spectra. For other Ap stars, its presence can easily be mistaken for Fe~{\sc i}~6335.330\,{\AA} ($\log gf=-2.177$, $\chi_{\rm ex}=2.1979$\,eV), but for {\cpd}, the observed central wavelength, Zeeman pattern, and symmetry of the feature all indicate that the contribution from Fe~{\sc i} is negligible. Similar to the case of Th~{\sc iii}~6599\,{\AA}, this can also be confirmed through inspection of Fe~{\sc i} lines that should be even stronger than the 6335.330\,{\AA} line. 

The 6243\,{\AA} line (see Fig.~\ref{fig:oprofsth3}) is another of the most obvious Th~{\sc iii} features, and for {\cpd}, it appears to be a rare example (regardless of ion) of a mostly blend-free line during both the {\blo} and {\bhi} phases. Another that fits in the same category and that was also exquisitely split during all of the UVES observations is Th~{\sc iii}~8105.394\,{\AA} line. The region around this line can be mildly contaminated by telluric absorption under poor observations conditions, but our linelist includes no likely stellar absorption contaminants in the vicinity. The results displayed in Figure~\ref{fig:prot} can in fact be closely duplicated by applying the same analysis described in Section~\ref{analysis} to Th~{\sc iii}~8105\,{\AA}. This is also true for Th~{\sc iii} 6243, 7461, and 7823\,{\AA}. Most of the other Th~{\sc iii} lines become blended with neighboring during {\bhi} phases due to increased widths of the RMSL. 

There are a few more details worth mentioning before moving on. The Th~{\sc iii} lines at 6881, 7243, 7921, and 9053\,{\AA} (see Fig.~\ref{fig:oprofsth3}) all fall in regions of likely telluric contamination. The 7921\,{\AA} line is the most magnetically insensitive of the displayed Th~{\sc iii} lines, and although a weak telluric feature is almost certainly blended with the observed feature, the lack of RMSL nonetheless supports the identification of the line as Th~{\sc iii}.


\begin{figure*}
\includegraphics[width=\textwidth]{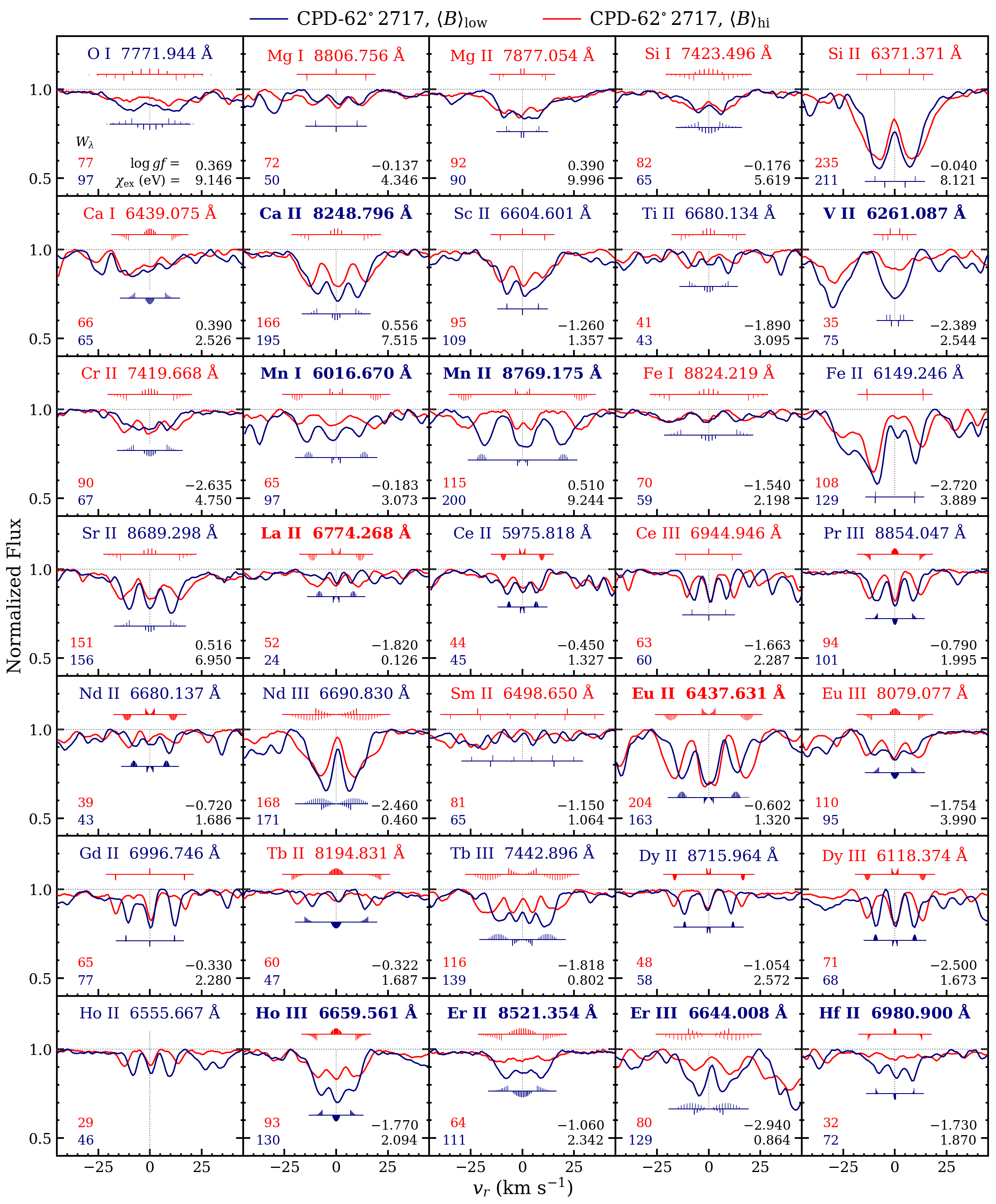}
\caption{The periodic table of peculiar at $\lambda>5700$\,{\AA} in the UVES spectra of {\cpd}. Meanings are mostly the same as in Figure~\ref{fig:oprofsth3}. With the exception of Hf~{\sc ii}, for which only one line seems to be present, each panel shows a representative line from an ion whose presence is supported by the central positions, relative intensities, and Zeeman patterns of $>1$ and often numerous lines. The color coding of the ion \& wavelength labels corresponds to the coloring of the spectra, such that for example if the line is strongest in the {\bhi} spectrum, the label is red. Bold font indicates cases of significant variability whereby the line strength changes by $>25$\,m{\AA} between the {\blo} and {\bhi} phases. Note that full atomic data is not available for the Ho~{\sc ii} line \citep[rest wavelength taken from][]{2019ApJS..240...28O} and the quoted $W_{\lambda}$ were measured manually. \label{fig:oprofs}}
\end{figure*}

\section{Other Elements in the Optical} \label{opticalions}
\subsection{Light Elements} \label{lightelements}
Figure~\ref{fig:oprofs} shows profiles of lines from 35 ions that are likely or definitely present in the 5700--9000\,{\AA} range in the UVES spectra of {\cpd}. The identifications are based on reasonable matches between the observed and expected central positions, relative intensities, and Zeeman patterns of numerous lines in most cases. 

The lightest metal lines whose presence can be confirmed are the O~{\sc i}  7772, 7774, and 7775\,{\AA} triplet, and although quite weak and blended in the case of O~{\sc i} 7775\,{\AA}, the identifications can be confirmed by the widely split 7772\,{\AA} and 7774\,{\AA} lines. Mg, Si, Ca, Sc, and Ti are the next lightest elements present, albeit represented by just a few lines each. With the exceptions of Sc and Ti, for which lines from the neutral variety are not detected, the singly-ionized variety is far stronger than the neutral. Si~{\sc ii} and Ca~{\sc ii} in particular are likely to be overabundant on the surface of {\cpd} based on comparison to archival UVES spectra of cool Ap stars. By that same argument, Ti~{\sc ii} may be close to solar if not slightly below.

As for the iron peak elements, the trends established in the $H$-band spectra of {\cpd} are repeated in the optical. Namely, lines from V~{\sc ii} and Mn~{\sc ii} are abnormally strong during the {\blo} phases. This seems to be a quite rare abundance anomaly. We've been unable to identify any other Ap star whose archival UVES spectra shows stronger V~{\sc ii} lines than those of {\cpd}. The lines are simply not present in the spectra of most Ap stars. Mn~{\sc ii} lines on the other hand are often present, just not at the strengths observed in the {\blo} phase of {\cpd}. 

Note that the identification of the RMSL around 6016.7\,{\AA} as Mn~{\sc i} is tentative. In addition to the positions of the magnetically split components being slightly offset from expectations (possibly due to blending), a few other Mn~{\sc i} lines (6013.510\,{\AA} and 6021.820\,{\AA}) that should be of comparable strength do not seem to be present. Further, whereas the variability trend of the possible Mn~{\sc i}~6016.670\,{\AA} line follows that of the Mn~{\sc ii} lines, that does not necessarily confirm the identification. We can only identify just over half of the lines in a given wavelength range, meaning that at least one (more likely several) important ion is missing from our linelist. 

Similar to PS, unblended Fe lines are not easy to find in the spectra of {\cpd}. Considering the long-running debate over whether Fe was missing from the atmosphere of PS \citep{przybylski1977}, we suspect that Przybylski would be amused to see another example of this, though with a far stronger magnetic field in this case. Fe is certainly present of course, and based again on comparison to archival UVES spectra of other Ap stars for which abundances have been estimated, Fe~{\sc i} may be slightly underabundant with respect to Solar, while Fe~{\sc ii} may be slightly overabundant. Whatever the case, one thing is certain: RMSL from heavier elements are far more obvious in the 5700--9000\,{\AA} region.

Sr~{\sc ii} is the only ion between the iron peak and the REE that we've been able to confirm the presence of in the {\cpd} spectra, and it appears to be severely enhanced as is empirically expected for Ap stars in this temperature regime. The Sr~{\sc ii} resonance lines at 4077.709\,{\AA} and 4215.519\,{\AA} are among the strongest metal features in the blue UVES spectra, and another Sr~{\sc ii} line at 4305.443\,{\AA} is also present and quite strong. As for the 5700--9000\,{\AA} region, the Sr~{\sc ii} lines at 8506.006\,{\AA} ($\chi_{\rm ex}=6.915$\,eV, $\log gf=0.270$, $g_{\rm eff}=0.83$) and 8689.298\,{\AA} ($\chi_{\rm ex}=6.950$\,eV, $\log gf=0.516$, $g_{\rm eff}=1.10$) are expected to be strongest and should be magnetically split in the case of {\cpd}. Indeed, these lines appear to be present with the expected flux ratios ($W_{\lambda}$ of Sr~{\sc ii} 8689\,{\AA} is about twice that of Sr~{\sc ii}~8506\,{\AA}) and also with the expected Zeeman patterns (quasi-triplets). However, unlike most of the other ions represented in Figure~\ref{fig:oprofs}, we found that it was necessary to apply a $14.37$\,km\,s$^{-1}$ velocity correction to the rest wavelengths quoted by VALD in order for the central components of the observed features to align with $v_{\rm r}=0$ (as in Figure~\ref{fig:oprofs} for Sr~{\sc ii} 8689\,{\AA}). The cause of these wavelength discrepancies is unclear.

Although Y~{\sc ii} and Ba~{\sc ii} are often present in the spectra of cool Ap stars, with Ba~{\sc ii} often accounting for several of the strongest metal lines in the optical, the spectra of {\cpd} show little to no evidence of these ions.

\subsection{Rare Earth Elements}
Most of the REE are definitely present in the {\cpd} spectra, albeit with the light REE2 being quite weak. In particular, only a small handful of unblended RMSL from La~{\sc ii}, Ce~{\sc ii}, and Sm~{\sc ii} can be found between 5700--9000\,{\AA}, while Pr~{\sc ii} lines are exceedingly weak if not absent altogether. Nd~{\sc ii} is far better represented and also more easily confirmed thanks to several relatively magnetically insensitive lines (e.g., Nd~{\sc ii}~6580.930\,{\AA}, with $g_{\rm eff}=0.23$). Eu~{\sc ii} is almost always strong in the spectra of cool Ap stars, and the situation is no different for {\cpd}. However, the strengths of the Eu~{\sc iii} lines (10 of which appear to be present) are particularly noteworthy since the same lines are absent from the spectra of PS and from most Ap stars in general.

It is worth noting that the observed positions of Sm~{\sc ii} and Eu~{\sc iii} lines in the {\cpd} spectra are systematically offset from the VALD wavelengths by about $3.5$\,km\,s$^{-1}$, and the velocities of the associated line in Figure~\ref{fig:oprofs} have been corrected by that value. Unlike the situation with Sr~{\sc ii} (see Section~\ref{lightelements}), this may be due to Sm~{\sc ii} and Eu~{\sc iii} overabundance spots on the surface of {\cpd} whose locations strongly differ from those of the Pr and Dy lines used for radial velocity measurement.

Along with Th~{\sc iii}, it is the contributions from heavy REE3 that set {\cpd} apart from other Ap stars. Lines from Dy~{\sc ii} are numerous in the {\cpd}, with at least 83 lines present and exquisitely split like the 8715.964\,{\AA} line shown in Figure~\ref{fig:oprofs}. Of all the ions present, Dy~{\sc ii} is actually the easiest to confirm, far easier than Fe for example. Other ions that fall into this category include Pr~{\sc iii} (at least 43 RMSL), Gd~{\sc ii} (at least 60 RMSL), and Tb~{\sc iii} (at least 43 RMSL).

The Ho and Er lines of {\cpd} are also noteworthy considering that we are aware of no other Ap star for which they stronger. The only stars that we know of which even come close are the cool Ap stars HD\,81588 and HD\,143487 (which will be discussed in a subsequent section of this paper). Unfortunately, the Ho~{\sc ii}, Ho~{\sc iii}, and Er~{\sc iii} linelists are quite limited, and some parts of the Er~{\sc iii} linelist may in fact be erroneous given lines like the one at 5903.284\,{\AA}, for which the observed and expected splitting patterns and splitting widths are a severe mismatch. 

\begin{figure*}
\includegraphics[width=\textwidth]{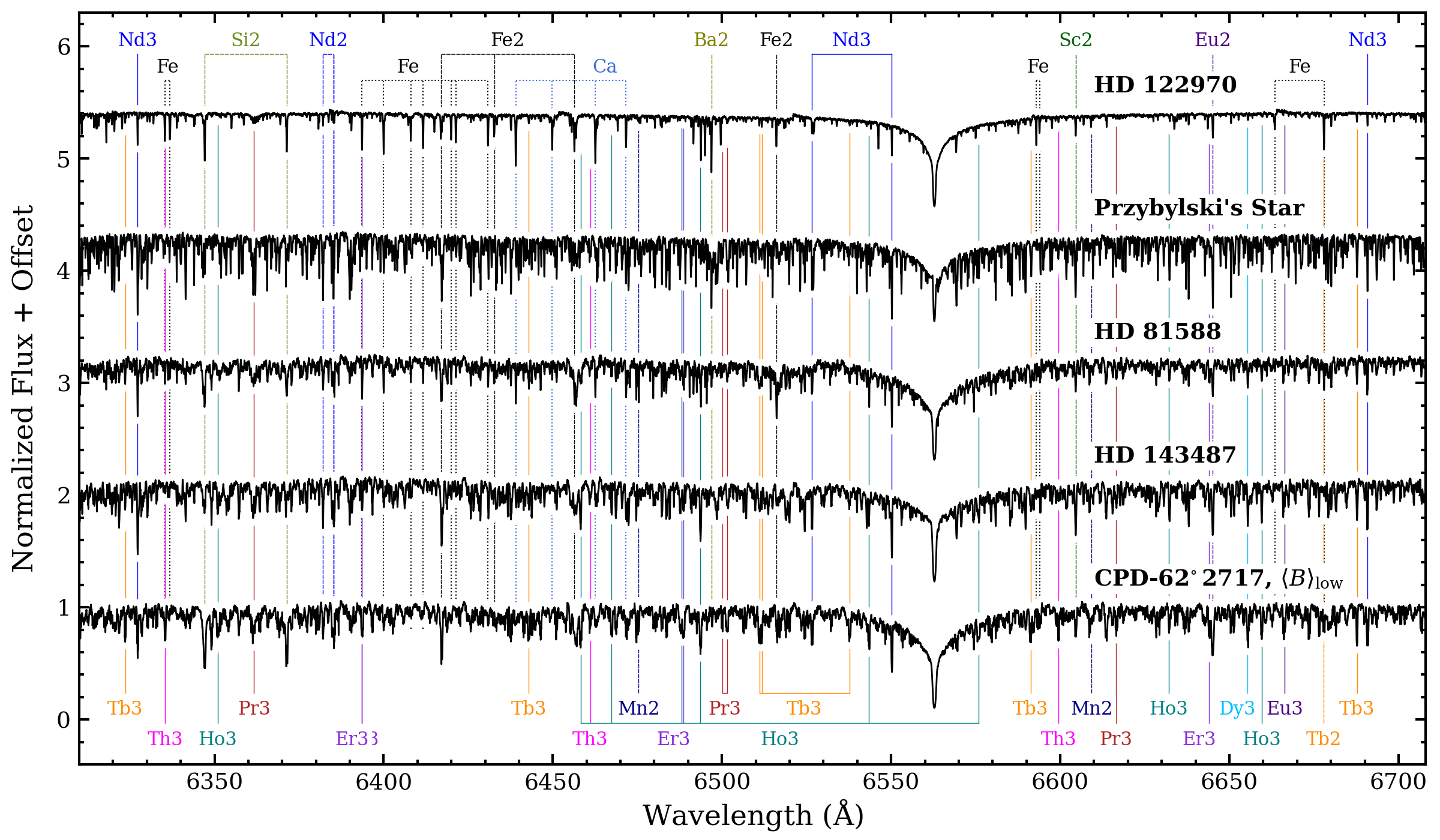}
\caption{Comparison of the mean UVES {\blo} spectrum of {\cpd} to archival UVES spectra of other Ap stars that we identified as possibly exhibiting the Th~{\sc iii} lines. \label{fig:optical}}
\end{figure*}

Although VALD currently lacks atomic data for Ho~{\sc ii} lines at wavelengths longer than 4212\,{\AA}, \citet{2019ApJS..240...28O} classified numerous lines up to 7000\,{\AA} as  Ho~{\sc ii} and Ho~{\sc iii}. The most isolated and clearly split of the Ho~{\sc ii} lines (6555.667\,{\AA}) from that work is shown in Figure~\ref{fig:oprofs}. There are a few nearby lines that might show up in stellar spectra (Pr~{\sc ii}~6555.609\,{\AA}, Ce~{\sc i}~6555.649\,{\AA}, and Mn~{\sc ii}~6555.686\,{\AA}), but those lines are not expected to be present in the {\cpd} spectra given the weakness or non-detections of stronger lines of the associated ions. Further, the observed Zeeman pattern empirically indicates $g_{\rm eff}\sim1.2$, which is inconsistent with the other possible contaminants. Some of the other Ho~{\sc ii} lines from \citet{2019ApJS..240...28O} that appear to be present in the $>5700$\,{\AA} region of the {\cpd} spectra include 6985.076\,{\AA} (doublet pattern), 6976.756\,{\AA} (triplet), 6655.714\,{\AA} (triplet), 6208.607\,{\AA} (triplet), 6073.607\,{\AA} (triplet), 6060.207\,{\AA} (triplet), 6058.185\,{\AA} (triplet), 6022.252\,{\AA} (triplet), 6005.267\,{\AA} (doublet), 5993.057\,{\AA} (triplet), 5961.704\,{\AA} (triplet), 5961.195\,{\AA} (low-$g_{\rm eff}$?), 5904.277\,{\AA} (triplet), 5866.437\,{\AA} (triplet), and 5839.441\,{\AA} (triplet). Most of these lines are blended to some extent. 

Surprisingly, we have found no evidence for lines from the heaviest REE -- Tm, Yb, and Lu -- in the {\cpd} spectra, or at least none that are not buried beneath a plethora of stronger lines. 

\subsection{Heavy Elements}
The only element between the REE and thorium that seems to possibly be present is Hf~{\sc ii}, which immediately follows the REE in the periodic table at $Z=72$. There are depressions in the continuum at the positions of most of the Hf~{\sc ii} lines with $\log gf>-2$ in the 5700--9000\,{\AA} region, but the nearly all of these suffer from blending. The Hf~{\sc ii} 6980.900\,{\AA} line is the best example we could find and it too appears to be blended, with possible telluric contributions. However, our linelist includes no other likely candidates in the vicinity that would contribute to the observed feature. Furthermore, the variability of the possible Hf~{\sc ii} line is quite similar to that of the heavy REE and Th~{\sc iii}, such that Hf~{\sc ii} may indeed be the primary contributor to the 6980.900\,{\AA} RMSL. 

The weakness if not lack of Th~{\sc ii} features in the {\cpd} spectra is somewhat surprising. Much like Hf~{\sc ii}, there are depressions in the continuum at the positions of most of the Th~{\sc ii} lines that are expected to be strongest, but only a few of them exhibit Zeeman patterns consistent with expectations. The essential non-detection of Th~{\sc ii} does not necessarily hinder the Th~{\sc iii} line identifications however. Praseodymium exhibits similar behavior, whereby lines from Pr~{\sc ii} are weak/absent while lines from Pr~{\sc iii} are among the strongest in the spectra.

This section would not be complete without mentioning uranium. Although it is expected be present in stellar atmospheres and is likely be enhanced in the atmosphere of {\cpd} given the thorium enhancement, we could find no solid evidence for U~{\sc ii} lines in the spectra. The same argument from the last paragraph applies equally well here; if anything, we would expect to see lines of U~{\sc iii}. It remains to be seen if some of the unknown lines in the optical spectra of {\cpd} can be attributed to U~{\sc iii}, and although it is understandably a tall order, the need for a laboratory-determined U~{\sc iii} linelist covering the entire optical wavelength regime has never been more urgent.

\begin{figure*}
\includegraphics[width=\textwidth]{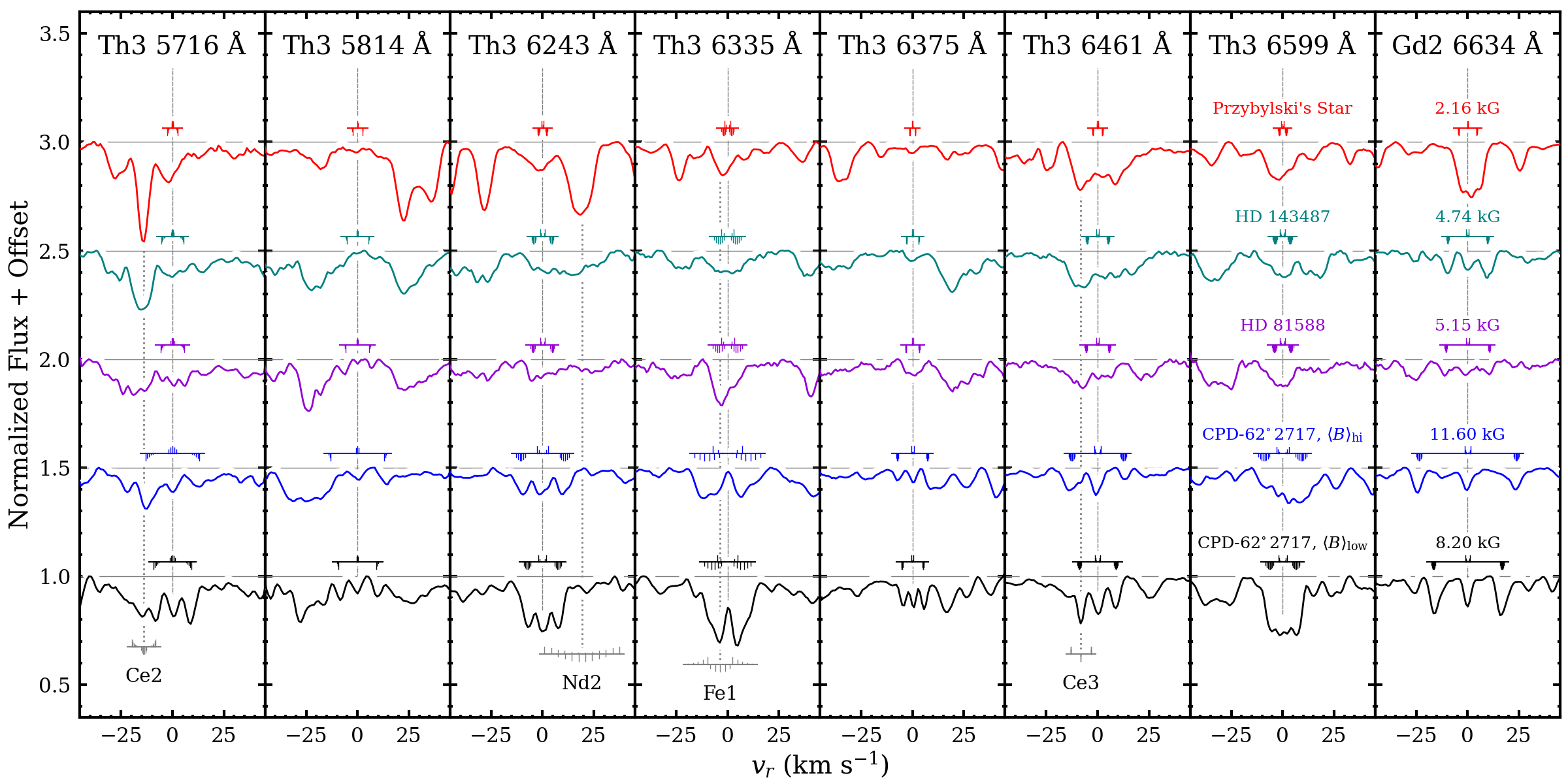}
\caption{Comparison of Th~{\sc iii} line profiles of {\cpd}, HD\,81588, HD\,143487, and Przybylski's Star. The magnetically sensitive ($g_{\rm eff}=2.2$) Gd~{\sc ii} 6634\,{\AA} line is also shown for context, since its magnetically split components are resolved in the spectra of all four stars. \label{fig:compTh3}}
\end{figure*}

\section{Is the Thorium Peculiarity Unique to {\cpd}?} \label{compareth}
Given the nature of the discovery at hand, the expectation that it is unlikely to be unique, and our failure to find further examples among the APOGEE Ap star sample, it was only prudent to search the ESO spectra archive for additional stars with Th~{\sc iii} lines. In hopes of this search being aided by Zeeman coincidence statistics, we started by inspecting the UVES and/or HARPS spectra available for 57 of the 84 Ap/Bp stars known to exhibit the magnetically split components of the Fe~{\sc ii}~6149\,{\AA} line \citep{2003A&A...403..645B, 2005A&A...440L..37H, freyhammer2008, 2010MNRAS.402.1883E, 2012MNRAS.420.2727E, mathys2017}. We also checked the archival spectra of numerous Ap/Bp stars that have been identified as particularly peculiar in the literature despite lack of RMSL. Considering the Gaia DR3 $T_{\rm eff}<6834$\,K of {\cpd}, we also included in our search the spectra of numerous stars for which $T_{\rm eff}<7500$\,K has been estimated. 

The quest for {\cpd} analogues was ultimately narrowed to just three stars -- PS, HD\,81588, and HD\,143487. Archival UVES spectra of these stars are displayed in Figure~\ref{fig:optical} along with the mean {\blo} UVES spectrum of {\cpd} and a UVES spectrum of the relatively pedestrian roAp star HD\,122970. The latter is included simply emphasize the extreme peculiarity of the others. but HD\,122970 is also one of the few Ap stars for which a thorium abundance has been estimated; \citet{ryabchikova2000} reported an upper limit on [Th/H] of $+$2.11 dex with respect to solar based on analysis of four Th~{\sc ii} lines. We see no strong evidence for thorium lines in the spectra of this star.

All of the spectra displayed in Figure~\ref{fig:optical} have the same $R=107\,000$ as those of {\cpd}, and the spectra of HD\,122970 and PS also cover the same wavelength range (5655--9464\,{\AA}) as those of {\cpd}. The wavelength coverage (4959--7071\,{\AA}) of the HD\,81588 and HD\,143487 spectra unfortunately does not extent as far to the red and thus does not cover isolated Th~{\sc iii} features such as the 8105\,{\AA} line. The observation of HD\,122970 took place on 2008 August 3 and achieved S/N$=168$ in a 500\,s exposure. Numerous ESO archive spectra are available for PS, but since we didn't see much evidence of temporal variability, we simply make use of the UVES observation from 2006 January 14, which achieved S/N$=267$ in a 680\,s exposure. The only UVES of observation of HD\,81588 took place on 2008 February 19 and achieved S/N$=330$ in a 2200\,s exposure. Four UVES spectra from 2007--2010 are available for HD\,143487, and due to the associated lack of significant spectroscopic variability, we simply make use of the highest-S/N spectrum that was obtained on 2010 June 27 and that achieved S/N$=325$ in a 3720\,s exposure. 

We could not help but notice that the four archival UVES spectra of HD\,143487 show $>10$\,km\,s$^{-1}$ radial velocity variations, with the 2007 and 2010 observations having $v_{r}\sim-25$\,km\,s$^{-1}$ and with the two 2008 observations having $v_{r}\sim-16$\,km\,s$^{-1}$, thus implying that it is a member of a binary system. This was also noticed by the Gaia collaboration\footnote{\href{https://vizier.cds.unistra.fr/viz-bin/VizieR-5?-ref=VIZ6425b1701e89e1&-out.add=.&-source=I/357/tboasb1c\&recno=28762&-out.orig=o}{Gaia DR3 orbital solution for HD\,143487}}, who found it to be single-lined spectroscopic binary (SB1) with a long 1354 day orbital period. Likewise, the UVES spectra show no evidence of a second set of lines.

In Figure~\ref{fig:optical}, lines that are strong for typical cool Ap stars like HD\,122970 are labeled along the top. With just a few exceptions, those same features are either absent, weak, or unrecognizable due to blending for the other stars. PS is well known as the most peculiar Ap star, and one does not need be an expert in this genre to recognize why from a glance at Figure~\ref{fig:optical}. Only a few of the strong lines of PS are labeled, but the majority of the unlabeled ones can be attributed to light REE2 (with particularly numerous contributions from Ce~{\sc ii}, Nd~{\sc ii}, and Sm~{\sc ii}), light REE3, or to ions that are missing from existing linelists. Given that blank continuum regions in optical spectra of PS are rare (if not non-existent), it was inevitable that PS would be included in this discussion. There are simply local minima everywhere, and the presence of Th~{\sc iii} (and just about any ion for that matter) cannot be ruled out. 

Given the sparse literature mentions of HD\,81588 and HD\,143487, the degree of peculiarity of these stars came as somewhat of a surprise. Both stars are close spectroscopic analogues of {\cpd}, and they are in fact the only examples we could find of Ho~{\sc iii} and Er~{\sc iii} line strengths even remotely close to those of {\cpd}. All three stars are more extreme than PS in that respect.

The only study we are aware of that specifically discussed HD\,81588 was \citet{2012MNRAS.420.2727E}, and these authors noted that the star ``showed a highly peculiar spectrum with strong rare earth element lines of Nd~{\sc iii} and Pr~{\sc iii},''. They also estimated $T_{\rm eff}=7400$\,K, thus putting HD\,81588 near the extreme cool end of the Ap star temperature distribution. The Fe~{\sc ii}~6149\,{\AA} line was found to be magnetically split with an implied {\bmag}$=2.4\pm0.2$\,kG. We confirmed this to be the case in the UVES spectrum as well, but we also found that the {\bmag} implied by virtually all of the other RMSL (most of which are REE lines) was more than double that of Fe~{\sc ii} at $\sim5.2$\,kG. This very large discrepancy is likely caused by significantly highly inhomogeneous distributions of Fe and REE on the surface of HD\,81588, which results in a different weighting of the local field intensities in the averaging over the visible stellar disk. 

HD\,143487 was first specifically mentioned in the literature by \citet{freyhammer2008}, and in this case, splitting of the Fe~{\sc ii}~6149\,{\AA} line indicated {\bmag}$=4.23\pm0.07$\,kG. These authors also reported $T_{\rm eff}=6930$\,K for HD\,143487, thus also placing it toward the lower limit of $T_{\rm eff}$ for Ap stars and making it comparable to {\cpd}, PS, and HD\,81588. Similar to the case of HD\,81588, \citet{freyhammer2008} also pointed out that the ``spectrum of HD\,143487 is highly peculiar and, e.g. Pr, Nd are among the strongest of this study.'' \citet{elkin2010} subsequently discovered that HD\,143487 is a roAp star with an 8.8 minute pulsation period, and this period was then refined to 9.63 minutes by \citet{2013MNRAS.431.2808K}.

The two strong Th~{\sc iii} lines at 6335\,{\AA} and 6599\,{\AA} are labeled along the bottom of Figure~\ref{fig:optical}, and only for {\cpd} are they obvious. In order to check for their presence in the other stars, it is necessary to view the line profiles in detail. Figure~\ref{fig:compTh3} does just that, showing six of the Th~{\sc iii} lines that are covered by the spectra of all four stars and that are clearly split in the {\cpd} spectra. The Gd~{\sc ii} 6634\,{\AA} line is also included for context, since its magnetically split components are resolved at $R=107\,000$ for all four stars. In the large upper panels, the Zeeman patterns above each line profile have been scaled horizontally according to the {\bmag} values given in the Gd~{\sc ii} panel at far right. For HD\,81588, the {\bmag} implied by REE lines has been adopted. When it was possible to identify blended features (usually it was not possible; e.g., the features right of Th~{\sc iii}~6375\,{\AA} and left of Th~{\sc iii}~6599\,{\AA}), the associated Zeeman patterns are shown along the bottom of the larger panels of Figure~\ref{fig:compTh3}, scaled for {\bmag}$=8.2$\,kG. 

For PS, the Th~{\sc iii} lines all appear to be partial blends; only for the two weakest lines (5814\,{\AA} and 6375\,{\AA}) do local minima coincide closely with the Th~{\sc iii} wavelengths. The evidence for Th~{\sc iii} is slightly better for HD\,143487 considering that a local minimum coincides with Th~{\sc iii}~6599\,{\AA} and also that the blue triplet component of Th~{\sc iii}~6243\,{\AA} appears possibly to be resolved. 

HD\,81588 on the other hand shows quite convincing evidence of Th~{\sc iii}. The magnetic splitting of the 5716\,{\AA} and 5814\,{\AA} lines appears to be resolved, and we are unaware of any other possible identifications beyond Th~{\sc iii}. Similar to HD\,143487, the blue triplet component of Th~{\sc iii}~6243\,{\AA} also appears to be resolved for HD\,81588. The Th~{\sc iii}~6243\,{\AA} line is unfortunately contaminated by a nearby Fe~{\sc i} line, but the position of the extended red wing of the observed feature is consistent with a contribution from Th~{\sc iii}. A hint of magnetic splitting can be seen in Th~{\sc iii}~6375\,{\AA}, albeit not fully resolved. Th~{\sc iii}~6461\,{\AA} is a blend for all four stars, with the primary contaminant usually being a nearby Ce~{\sc iii} line. However, only for {\cpd} and HD\,81588 does the Th~{\sc iii}~6461\,{\AA} rest wavelength coincide with a local minimum.

Although we consider our search for additional stars with Th~{\sc iii} RMSL inconclusive for the time being, it is important to reiterate the point that was made earlier on (see Figures~\ref{fig:hprofsth3} and \ref{fig:oprofsth3}). Namely, whereas the Th~{\sc iii} features of {\cpd} are blatantly present and strong during {\blo} phases, they are far less obvious during {\bhi} phases. The situation could very well be similar for HD\,143487 and HD\,81588. For neither star is the amplitude of {\bmag} variation nor rotation period known, and thus one cannot rule out the possibility that future observations will take place when severely Th-enhanced portions of their surfaces are in the line-of-sight. We therefore recommend long-term spectroscopic monitoring of HD\,143487 and HD\,81588 in order to constrain their rotation periods and check for variability similar to that of {\cpd}. Any such observations would ideally extend to $\sim$8400\,{\AA} and thus cover some of the isolated Th~{\sc iii} lines that would be expected show resolved magnetic splitting. 

Whereas Przybylski's Star remains uniquely peculiar, the arguably more extreme peculiarities of {\cpd} firmly place it among the most peculiar stars known. However, it remains to be seen how overabundant thorium actually is on the surface of {\cpd}, by what amplitude that abundance varies as a function of rotation phase, and whether or not any other stars can be shown to have comparable or even higher thorium abundances. Other worthwhile pursuits might include additional spectroscopic and spectropolarimetric monitoring in order to better constrain the rotation period and magnetic field geometry, spectroscopy in other wavelength regimes to search for additional heavy elements, and high-cadence spectroscopy or photometry to check for pulsation. We eagerly anticipate future efforts to address these open questions.

\section*{Acknowledgements}

Funding for the Sloan Digital Sky Survey IV has been provided by the Alfred P. Sloan Foundation, the U.S. Department of Energy Office of Science, and the Participating Institutions. SDSS acknowledges support and resources from the Center for High-Performance Computing at the University of Utah. The SDSS web site is www.sdss.org.

SDSS is managed by the Astrophysical Research Consortium for the Participating Institutions of the SDSS Collaboration including the Brazilian Participation Group, the Carnegie Institution for Science, Carnegie Mellon University, the Chilean Participation Group, the French Participation Group, Harvard-Smithsonian Center for Astrophysics, Instituto de Astrof\'{i}sica de Canarias, The Johns Hopkins University, Kavli Institute for the Physics and Mathematics of the Universe (IPMU) / University of Tokyo, Lawrence Berkeley National Laboratory, Leibniz Institut f\"{u}r Astrophysik Potsdam (AIP), Max-Planck-Institut f\"{u}r Astronomie (MPIA Heidelberg), Max-Planck-Institut für Astrophysik (MPA Garching), Max-Planck-Institut f\"{u}r Extraterrestrische Physik (MPE), National Astronomical Observatories of China, New Mexico State University, New York University, University of Notre Dame, Observat\'{o}rio Nacional / MCTI, The Ohio State University, Pennsylvania State University, Shanghai Astronomical Observatory, United Kingdom Participation Group, Universidad Nacional Aut\'{o}noma de México, University of Arizona, University of Colorado Boulder, University of Oxford, University of Portsmouth, University of Utah, University of Virginia, University of Washington, University of Wisconsin, Vanderbilt University, and Yale University.

This work has made use of data from the European Space Agency (ESA) mission Gaia (https://www.cosmos.esa.int/gaia), processed by the Gaia Data Processing and Analysis Consortium (DPAC, https://www.cosmos.esa.int/web/gaia/dpac/consortium). Funding for the DPAC has been provided by national institutions, in particular the institutions participating in the Gaia Multilateral Agreement.

Based on observations made with ESO telescopes at the La Silla Paranal Observatory under programme IDs 0103.D-0119(A), 0105.D-0193(A), and 0109.D-0226(A).

\section*{Data Availability}

With the exception of the 2023 April 27 SDSS/APOGEE spectrum of {\cpd}, which is proprietary until the next SDSS data release, all of the \href{https://dr17.sdss.org/infrared/spectrum/search}{SDSS/APOGEE} and \href{http://archive.eso.org/scienceportal/home}{ESO/VLT} data used in this paper are publicly available.



\bibliographystyle{mnras}
\bibliography{ms} 




\bsp	
\label{lastpage}
\end{document}